\documentclass[12pt]{article}

\usepackage{color}
\usepackage{amsmath,amssymb}
\usepackage[dvips]{graphicx,psfrag}
\usepackage{latexsym}
\usepackage{cite}
\usepackage{wrapfig}
\usepackage{pstricks}
\usepackage{eepic}
\usepackage{epic}

\allowdisplaybreaks[1]

\newcommand{\eq}[1]{(\ref{#1})}
\newcommand{\nn}{\nonumber}
\newcommand{\ds}{\displaystyle}

\newcommand{\vev}[1]{\left\langle #1 \right\rangle}

\newcommand{\del}{\partial}

\newcommand{\asymeq}{\underset{asym}{\simeq}}

\newcommand{\bP}{\boldsymbol{P}}
\newcommand{\bQ}{\boldsymbol{Q}}

\DeclareMathOperator{\tr}{tr}

\DeclareMathOperator{\diag}{diag}

\newtheorem{Theorem}{Theorem}

\renewcommand{\thefootnote}{\fnsymbol{footnote}}

\makeatletter
\@addtoreset{equation}{section}

\begin{document}


\begin{titlepage}
\thispagestyle{empty} 
\begin{flushright}
September 2011
\end{flushright}

\vspace{2.3cm}

\begin{center}
\noindent{\Large \textbf{
Stokes Phenomena and Quantum Integrability\\
\vspace{0.3cm} in Non-critical String/M Theory
}}
\end{center}

\vspace{1cm}

\begin{center}
\noindent{Chuan-Tsung Chan\footnote{ctchan@thu.edu.tw}$^{,p}$, 
Hirotaka Irie\footnote{irie@phys.cts.nthu.edu.tw}$^{,q}$
and Chi-Hsien Yeh\footnote{d95222008@ntu.edu.tw}$^{,r}$}
\end{center}
\vspace{0.5cm}
\begin{center}
{\it 
$^{p}$Department of Physics, Tunghai University, Taiwan, 40704\\
\vspace{.3cm}
$^{q}$National Center for Theoretical Sciences, \\
National Tsing-Hua University, Hsinchu 30013, Taiwan, R.O.C.\\
\vspace{.3cm}
$^{r}$Department of Physics and Center for Theoretical Sciences, \\
National Taiwan University, Taipei 10617, Taiwan, R.O.C. 
}
\end{center}

\vspace{1.0cm}

\begin{abstract}
We study Stokes phenomena 
of the $k\times k$ isomonodromy systems 
with an arbitrary Poincar\'e index $r$, especially 
which correspond to the fractional-superstring (or parafermionic-string) multi-critical points $(\hat p,\hat q)=(1,r-1)$ in the $k$-cut two-matrix models. 
Investigation of this system is important for the purpose of figuring out the non-critical version of M theory which was proposed to be the strong-coupling dual of fractional superstring theory 
as a two-matrix model with an infinite number of cuts. 
Surprisingly the multi-cut boundary-condition recursion equations have a universal form 
among the various multi-cut critical points, 
and this enables us to show explicit solutions of Stokes multipliers 
in quite wide classes of $(k,r)$. Although these critical points almost break 
the intrinsic $\mathbb Z_k$ symmetry of the multi-cut two-matrix models, 
this feature makes manifest a connection between 
the multi-cut boundary-condition recursion equations and the structures of quantum integrable systems. 
In particular, it is uncovered that the Stokes multipliers satisfy 
{\em multiple} Hirota equations (i.e.~{\em multiple} T-systems). 
Therefore our result provides a large extension of the ODE/IM correspondence to 
the general isomonodromy ODE systems endowed with the multi-cut boundary conditions. 
We also comment about a possibility that $\mathcal N=2$ QFT of Cecotti-Vafa 
would be ``topological series'' in non-critical M theory equipped with a single quantum integrability. 
\end{abstract}

\end{titlepage}

\newpage

\renewcommand{\thefootnote}{\arabic{footnote}}
\setcounter{footnote}{0}


\tableofcontents


\section{Introduction}

Non-perturbative aspects beyond perturbative string theory still remain
some of the missing pieces in our current understanding of string theory. 
In early study in the '90s, the following question about non-perturbative string theory was 
investigated: 
{\em what is the form of non-perturbative corrections to the perturbative expansions of string theory?}
Among various investigations, non-critical string theory \cite{Polyakov} played an important role since it is non-perturbatively formulated
by solvable matrix models \cite{DSL,TwoMatString,GrossMigdal2,    BIPZ,KazakovSeries,Kostov1,Kostov2,BMPNonP,BDSS,Kostov3,David0,MSS,David,DKK,TadaYamaguchiDouglas,DouglasGeneralizedKdV,Moore,GinspargZinnJustin,Shenker,fkn,DVV,EynardZinnJustin,fy12,fy3,Kris,
AnazawaIshikawaItoyama1,AnazawaIshikawaItoyama2,AnazawaItoyama,MultiCutUniversality,McGreevyVerlinde,Martinec,KMS,AKK,KazakovKostov,MMSS,SeSh2,fim,fi1,EynardLoop,EynardNPPartition,EynardMarino,CISY1,CIY1} and this fact makes it possible to see higher-order behavior of the string-theory free energy 
in strong string coupling constant $g_{\rm str}\to \infty$ \cite{GinspargZinnJustin,Shenker,EynardZinnJustin,David,fy12,fy3}. 
In particular, it was found that non-perturbative corrections to the perturbative amplitudes are given 
by order of $\mathcal O(e^{-*/g_{\rm str}})$, i.e.~open-string degree of freedom \cite{Shenker}: 
\begin{align}
\mathcal F(g_{\rm str}) = \ln \mathcal Z(g_{\rm str}) 
\asymeq \underbrace{\sum_{n=0}^\infty g_{\rm str}^{2n-2}\mathcal F_n}_{\text{\footnotesize perturb. part}} + \underbrace{\sum_I \theta_I \exp \Bigl[\sum_n g_{\rm str}^{2n-1}\mathcal F_n^{(I)}\Bigr] 
+ O(\theta_I^2)}_{\text{\footnotesize non-perturb. corrections}}, 
\label{PerturbExpansionIntro}
\end{align}
and this fact lead to a key idea of D-branes \cite{Polchinski}. 
The identification of D-branes was then confirmed in comparison between the Liouville-theory calculation \cite{Polyakov,KPZ,DDK,DOZZ,Teschner,FZZT,ZZ,sDOZZ,fuku-hoso,SeSh,KOPSS}, i.e.~ZZ-branes \cite{ZZ}, 
and succeeding calculations of matrix models \cite{McGreevyVerlinde,Martinec,KMS,AKK,KazakovKostov,SeSh2,fi1}. 
On the other hand, since they describe non-perturbative corrections in the perturbative asymptotic expansion, 
the relative weight factors $\bigl\{\theta_I\bigr\}_I$, called {\em D-instanton chemical potentials (fugacities)}, 
remain to be ambiguous parameters in perturbative string theory \cite{David}. 

\begin{figure}[t]
 \begin{center}
  \includegraphics[scale=0.6]{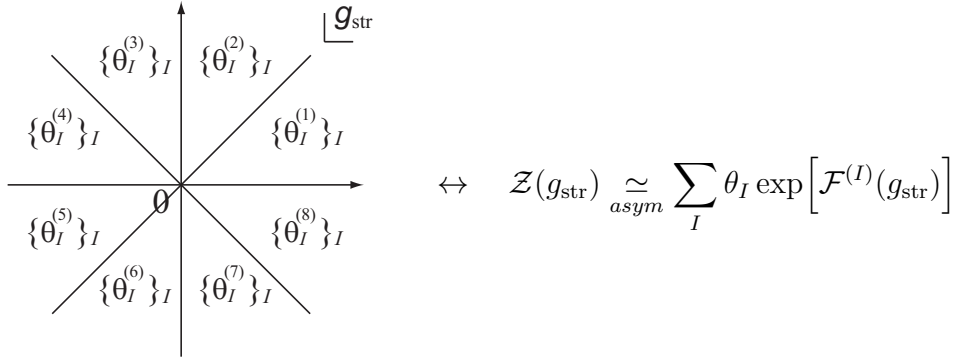} 
  \begin{picture}(190,0)(0,0)
  \put(20,62){$\ds \leftrightarrow\quad  \mathcal Z(g_{\rm str}) \asymeq \sum_I \theta_I \exp\Bigl[{\mathcal F^{(I)}(g_{\rm str})}\Bigr]$}
  \end{picture}
 \end{center}
 \caption{\footnotesize 
The asymptotic expansions can be different up to 
the D-instanton fugacities $\bigl\{\theta_I\bigr\}_I$ 
depending on from which direction of the string coupling $g_{\rm str}\,(\to 0 \times e^{i\chi})$ we expand the free energy (or partition function). 
That is, in this figure, there are eight different sets of the D-instanton fugacities $\bigl\{\theta_I^{(a)}\bigr\}_I\, 
(a=1,2,\cdots,8)$ 
depending on the angular domains of the string coupling $g_{\rm str}$. 
This mathematical property is called Stokes phenomenon. Therefore, determining Stokes phenomena means determining how the fugacities jump in analytic continuations crossing the domain wall 
(e.g.~$\theta_I^{(8)}\to \theta_I^{(1)}$ in crossing the positive real axes). 
Depending on the relative magnitude of perturbative free energy $\mathcal F^{(I)}(g_{\rm str})$ and the D-instanton fugacities, various perturbative string theories 
emerge as a dominant saddle point. This kind of phenomena can occur in any analytic continuations of 
the string-theory moduli spaces. 
} 
 \label{FigureStokes}
\end{figure}

A basic origin of the ambiguity is the fact that 
these parameters are subjected to discrete corrections in 
various analytic continuations of string-theory backgrounds 
and that the formal expansions of analytic equations can be performed 
without specifying the direction from which we expand the series (See Figure \ref{FigureStokes}). 
From the viewpoint of physics, this kind of phenomena make it possible to 
describe various different perturbative backgrounds in a single analytic theory. 
Mathematically these behaviors are known as {\em Stokes phenomena}, 
a basic analytic property in the asymptotic analysis. 
Determining the Stokes phenomena for the asymptotic expansions in the string coupling constant, 
Eq.~\eq{PerturbExpansionIntro}, 
(i.e.~how the fugacities $\bigl\{\theta_I\bigr\}_I$ jump in analytic continuations) 
can be performed by resurgent analysis of the perturbative asymptotic expansion Eq.~\eq{PerturbExpansionIntro}, 
which was analyzed recently \cite{GKIM,resurgentRef}. 
On the other hand, 
the non-perturbative ambiguities are information completely out of the asymptotic expansions, 
and therefore we have to come back to the original definition of matrix models and 
need to see {\em additional principle}
in order to determine the D-instanton fugacities in string theory. 
This is the question we investigate in this paper: 
{\em what is the principle to choose physical D-instanton fugacities?} 
A reason why we consider these non-perturbative problems so significant is because 
we believe that it is this principle which binds various string-theory vacua together to formulate
non-perturbatively completed string theory and consequently which guarantees the string dualities 
in a consistent way. Therefore, this would be the direction in which we can quantitatively 
tackle the string-theory landscape from the first principle. 

In study of this issue about the non-perturbative ambiguity, 
non-critical string theory is again expected to play an important rule. 
In non-critical string theory, the non-perturbative ambiguities are identified as 
integration constants of string equations \cite{David}. At the first sight, therefore it may 
seem plausible to say that 
the D-instanton fugacities are arbitrary free parameters in the non-perturbative theory. 
However, it is also true that there are several systems which prefer ``physical values'' rather than 
arbitrary values \cite{HHIKKMT}, 
and these physical values are implied in the matrix models. 
This kind of study has been investigated from various viewpoints \cite{FIK,
HHIKKMT,KKM,InstantonInCSFT,ChemicalPotentials1,ChemicalPotentials2,fis,ChemicalPotentials3,KurokiSugino,MSW1,MarinoHMSolution,MSW2,PasquettiSchiappa,MarinoLecture,CIY2}. 
In this paper, succeeding the previous study of the authors \cite{CIY2}, 
we investigate this issue in the framework of the multi-cut two-matrix models \cite{fi1,irie2}\cite{CISY1,CIY1,CIY2}. 

The multi-cut two-matrix models are the two-matrix version of the multi-cut one-matrix models \cite{MultiCut} 
and realize the multi-cut critical points which naturally extend the two-cut critical points of the matrix models 
\cite{GrossWitten,PeShe,DSS,Nappi,CDM,HMPN}. 
In particular, these critical points are controlled by multi-component KP hierarchy \cite{fi1} 
and this fact makes possible various quantitative analysis of the matrix models \cite{CISY1,CIY1,CIY2}. 
What is more, the two-cut critical points were found to describe 
minimal type 0 superstring theory \cite{TT,NewHat,UniCom}, 
and, as an extension of this, a special kind of the multi-cut critical points 
(which is also shown in Section \ref{SubSubSectionTwoCriticalPoints}) 
were proposed to describe minimal fractional superstring theory (or minimal parafermionic string theory) 
\cite{irie2}. An interesting new feature found by quantitative analyses of these critical points 
is the fact that these multi-cut systems generally include a number of perturbative vacua 
in its string-theory landscape \cite{CISY1} and also include various perturbative string-theory sectors 
in its wavefunctions \cite{CIY1}. 
In particular, the fractional-superstring critical points 
are found to describe a superposition of various minimal fractional superstring theories, 
and the chain of string theories 
is then interpreted as the extra dimension of the non-critical M theory \cite{CIY1}. 
This means that, in the case of an infinite number of cuts, 
there naturally appears a three-dimensional universal strong-coupling dual theory which 
realizes the philosophy proposed by Ho\v rava-Keeler for the non-critical M theory \cite{NonCriticalMTheory}. 
In this way, various new non-perturbative phenomena of string theory were 
revealed in the multi-cut critical points, and importantly these phenomena 
are fully described only after we determine the non-perturbative ambiguities, 
i.e.~they are governed by the information out of the perturbative description of string theory. 
Therefore, it is clear that the multi-cut matrix models provide a fruitful field in studying this issue,  
and the investigation of this system is expected to answer to the following question: {\em what is the non-perturbative completion of string theory?}

As a natural framework to study the ambiguity, 
we consider the isomonodromy description 
\cite{RHcite,RHPIIcite}\cite{Moore,FIK}\cite{ItsBook} for the multi-cut critical points.
Mathematically, the system is a $k\times k$ isomonodromy system 
with an irregular singularity of Poincar\'e index $r\, (=\hat q+1)$ which corresponds to 
the $k$-cut multi-critical points $(\hat p,\hat q)=(1,\hat q)$ in the multi-cut matrix models. 
In the isomonodromy study of matrix models, {\em the D-instanton chemical potentials are 
mapped to Stokes multipliers of the linear systems in the spectral parameter $\zeta$}
(i.e.~the Baker-Akhiezer systems Eq.~\eq{DESystem}) 
{\em through the Riemann-Hilbert procedure} (See \cite{ItsBook}). 
Roughly speaking, the Riemann-Hilbert procedure provides 
an integral expression of the string free energy $\mathcal F(g_{\rm str})$ which is given by 
the Stokes multipliers of the ODE systems. For instance in the case of the $2$-cut 
$(\hat p,\hat q)=(1,2)$ critical point, the RH integral is give 
by the Stokes multipliers $\{s_{n,i,j}\}_{n,i,j}$ of the corresponding ODE system as 
\begin{align}
\frac{\del^2 \mathcal F(t;g_{\rm str})}{\del t^2} = \bigl[f(t)\bigr]^2,
\qquad f(t)=\sum_{n} s_{n,2,1}   \int_{\mathcal K_n} \frac{d\lambda}{2\pi i} \,
 e^{\frac{1}{g_{\rm str}}\bigl(\varphi^{(2)}(t;\lambda)-\varphi^{(1)}(t;\lambda)\bigr)}+\cdots, \label{RHintegralIntroduction}
\end{align}
where $t$ is a parameter of the string-theory moduli space (i.e.~the worldsheet cosmological constant).%
\footnote{One can find an explanation of this integral in \cite{CIY2} and \cite{ItsBook}. }
In this way, the information of the non-perturbative ambiguities is 
translated into {\em the Stokes data of ordinary differential equation (ODE) systems} 
as the information which does not depend on the direction of asymptotic expansion 
(shown in Figure \ref{FigureStokes}).%
\footnote{For example in \cite{ItsBook}, by evaluating this integral, 
they show explicit expressions of how the figacities $\bigl\{ \theta_I \bigr\}_I$ 
(i.e.~$\bigl\{ \theta_I^{(a)} \bigr\}_I$ for each angular domain 
$(a=1,2,\cdots)$ in Figure \ref{FigureStokes}) can be expressed by 
the Stokes data $\{s_{n,i,j}\}_{n,i,j}$ of the ODE system of the Painlev\'e equations. }
Although the Stokes phenomena in these ODE systems are relatively simpler than those of non-linear differential 
equation systems like Painlev\'e equations, less is known in the case of large values of $k$ and $r$ 
compared with the isomonodromy systems of the second order Painlev\'e series \cite{ItsBook}. 

In the previous study of the authors \cite{CIY2}, the physical constraints (required from the matrix models) 
on the Stokes multipliers are formulated in this isomonodromy framework. 
As is reviewed in Section \ref{SubsubsectionMCgeometryAndStokesPhen}, the requirement is given as 
{\em the boundary conditions for physical cuts in the spectral curves}. 
Interestingly these physical constraints enable us to show explicit solutions of Stokes multipliers 
which satisfy high-degree algebraic equations called monodromy free (or cyclic) conditions 
(Eq.~\eq{OriginalMonodromyFreeCondition}), which are generally quite hard to 
solve without any help of the physical boundary conditions. 

In this paper, on the other hand, we extend this boundary condition 
to wider classes of the multi-cut critical points (general number of cuts, $k$, and general Poincar\'e index, $r$),
especially the critical points which correspond 
to minimal fractional superstring (or parafermionic string) theory \cite{irie2}. 
A major difference from the critical points of the previous study 
is lack of the $\mathbb Z_k$-symmetry which is intrinsic in the multi-cut two-matrix models. 
Therefore, the number of independent Stokes multipliers are $k$ times 
more than that of $\mathbb Z_k$ symmetric critical points, 
and this becomes a huge predicament to direct analysis of the Stokes phenomena. 
However, we will show that by this feature it becomes manifest 
that the multi-cut boundary conditions are equivalent to 
discrete Hirota equations (i.e.~T-systems) of quantum integrable systems. 
That is, the physical section of D-instanton chemical potentials are expressed by quantum integrability 
which would be an integrable structure of strong-coupling dual description, i.e.~non-critical M theory. 

\

Organization of this paper is as follows:
In Section \ref{SectionSummaryOfFSST}, basic facts about the fractional-superstring 
critical points are summarized. 
In Section \ref{SectionMCBC}, the multi-cut boundary conditions are studied. In particular, 
the T-systems of the Stokes multipliers are pointed out in Section \ref{SubsectionHirota}. Some explicit 
solutions are also shown in Section \ref{SubsectionExplicitSolutions}. 
Section \ref{SectionConclusionDiscussion} is devoted to conclusion and discussion. 
In Appendix \ref{SketchOfProofTheoremMCBCRE}, 
a brief proof of Theorem \ref{TheoremMultiCutBCRecursionEquations} is presented.

\section{Summary of fractional-superstring critical points \label{SectionSummaryOfFSST}}

In this section, we summarize basic facts about the fractional-superstring 
critical points \cite{irie2}, especially focusing on the results from the matrix-model analysis \cite{CISY1,CIY1,CIY2}, 
including the Stokes phenomena 
of the multi-cut matrix models which are generally developed in \cite{CIY2}. 

\subsection{Definition of the multi-cut matrix models}

The multi-cut two-matrix models are given by two-matrix models, 
\begin{align}
\mathcal Z = \int_{{\rm M}_N(\mathcal C^{(k)})\times {\rm M}_N(\mathcal C^{(k)})} dX dY\, e^{-N\tr \bigl[ V_1(X) + V_2(Y) -XY\bigr]},
\end{align}
with integration over the following special $N\times N$ normal matrices, ${\rm M}_N(\mathcal C^{(k)})$: 
\begin{align}
{\rm M}_N(\mathcal C^{(k)}) \equiv \bigl\{U \diag_{i=1}^k (\lambda_i) U^\dagger \bigl| U \in U(N), \, \lambda_i \in \mathcal C^{(k)} \subset \mathbb C\bigr\},
\end{align}
where the contour of the eigenvalues, $\mathcal C^{(k)}$, are defined as 
\begin{align}
\mathcal C^{(k)} 
= \bigcup_{n=0}^{k-1}\, \omega^n\, \mathbb R,
\qquad \omega= e^{2\pi i \frac{1}{k}},
\end{align}
in the case of the $k$-cut two-matrix models. 

An important ingredient of this system is the bi-orthonormal polynomial systems \cite{Mehta} 
defined by the following inner product, 
\begin{align}
\int_{\mathcal C^{(k)}\times \mathcal C^{(k)}} dx dy \, e^{-N\bigl[V_1(x)+V_2(y) -xy\bigr]} \, \alpha_n(x)\,\beta_m(y) = \delta_{n,m},
\end{align}
between the polynomials, 
\begin{align}
\alpha_n(x) = \frac{1}{\sqrt{h_n}}\Bigl(x^n + \cdots\Bigr),
\qquad \beta_n(y) = \frac{1}{\sqrt{h_n}}\Bigl(y^n + \cdots\Bigr). 
\end{align}
Here $\{h_n\}_{n\in \mathbb Z_+}$ are normalization coefficients. An interesting fact about the 
orthonormal-polynomial system is the following expression by the matrix integral \cite{GrossMigdal2}: 
\begin{align}
\alpha_n(x) = \frac{1}{\sqrt{h_n}}\bigl<\det \bigl(x-X\bigr)\bigr>_{\rm n\times n},\qquad 
\beta_n(x) = \frac{1}{\sqrt{h_n}}\bigl<\det \bigl(y-Y\bigr)\bigr>_{\rm n\times n}, 
\end{align}
where the expectation value $\bigl<\cdots \bigr>_{\rm n\times n}$ is given by $n\times n$ truncated 
matrices ${\rm M}_n(\mathcal C^{(k)})$: 
\begin{align}
\bigl<\mathcal O(X,Y)\bigr>_{\rm n\times n} \equiv 
\int_{{\rm M}_n(\mathcal C^{(k)})\times {\rm M}_n(\mathcal C^{(k)})} dX dY\, e^{-N\tr \bigl[ V_1(X) + V_2(Y) -XY\bigr]}\, \mathcal O(X,Y). \label{sizenMatrixIntegral}
\end{align}

Another important ingredient of this system is the resolvent operator $R(x)$ which 
contains the spectral information of the matrix integral: 
\begin{align}
R(x) =\frac{1}{N}\vev{\tr \frac{1}{x-X}} \quad \to\quad 
\int_{\mathcal C^{(k)}} d\lambda\, \frac{\rho (\lambda)}{x-\lambda}\qquad (N\to \infty). 
\label{DefinitionOfResolventAndEigenvalueDensityFunction}
\end{align}
In particular, the support of the eigenvalue density function $\rho(x)$ 
corresponds to discontinuity of the resolvent function: 
\begin{align}
\pi i \rho(\lambda) = R(\lambda + i \epsilon) - R(\lambda - i \epsilon), \qquad \lambda \in {\rm supp} \, (\rho) \subset \mathcal C^{(k)}. 
\end{align}
Since the orthonormal polynomials are also written with the resolvent operator: 
\begin{align}
\alpha_n(x) \sim \frac{1}{\sqrt{h_N}} \bigl<\det \bigl(x-X\bigr)\bigr>= \frac{1}{\sqrt{h_N}} \exp \Bigl[{\ds N \int^x d x' R(x')}\Bigr] \qquad (n\sim N), \label{ResolventOrthonormalPoly}
\end{align}
the above discontinuity should be observed also in the orthonormal polynomial of the large $N$ limit with $n/N\to 1$. 

\subsection{The ODE system and the weak-coupling results}

We move on to the continuum $(\hat p,\hat q)$ minimal fractional superstring theory \cite{irie2} 
by taking the scaling limit of the bi-orthonormal polynomials. The scaling limit is given by 
a lattice spacing $a$ as follows:
\begin{align}
\Psi_{\rm orth}(t;\zeta) \equiv 
\begin{pmatrix}
\psi_1(t;\zeta) \cr \vdots \cr \psi_k(t;\zeta)
\end{pmatrix},\qquad 
\psi_i (t;\zeta)= a^{\frac{\hat p}{2}} \omega^{\epsilon/2} \Bigl[(-1)^n \alpha_{kn+i-1}(x)\, e^{-NV_1(x)}\Bigr],
\label{ScalingOrthonormalPolynomials}
\end{align}
where the scaling valuables are introduced as 
\begin{align}
&x = a^{\frac{\hat p}{2}} \omega^{\epsilon/2}\zeta,\qquad
N^{-1} = g_{\rm str} \,a^{\frac{\hat p+\hat q}{2}} \to 0,\qquad 
\frac{kn}{N} = \exp\bigl(-t a^{\frac{\hat p+\hat q-1}{2}}\bigr) \to 1,\nn\\
&k^{-1}\del_n = - a^{1/2} \,g_{\rm str}\, \del_t \equiv - a^{1/2}\del  \to 0, 
\qquad(a\to 0). \label{ScalingArgument}
\end{align} 
The detail of these scaling limits is discussed in \cite{CISY1}. As will be mentioned later, 
there is a parameter $\epsilon\, (= 0,1)$, 
which are related to the two choices of fractional-superstring $(\hat p,\hat q)$ 
critical points. 
Consequently, the recursion equation of the orthonormal polynomial: 
\begin{align}
x \alpha_n(x) = \sum_{s\in \mathbb Z} A_s(n)\, \alpha_{n-s}(x),
\qquad N^{-1} \frac{\del}{\del x} \alpha_n(x) = \sum_{s\in \mathbb Z} B_s(n)\, \alpha_{n-s}(x)
\end{align}
becomes the following $k\times k$ differential equation system \cite{fi1}:
\begin{align}
\zeta \Psi(t;\zeta) = \bP(t;\del)\, \Psi(t;\zeta),\qquad 
g_{\rm str} \frac{\del}{\del \zeta} \Psi(t;\zeta)= \bQ(t;\del)\, \Psi(t;\zeta). \label{DESystem}
\end{align}
which is governed by $k$-component KP hierarchy \cite{kcKP}. 
Here we note that 
the scaling orthonormal polynomial $\Psi_{\rm orth}(t;\zeta)$ of Eq.~\eq{ScalingOrthonormalPolynomials} 
is a special solution to the Baker-Akhiezer function system Eqs.~\eq{DESystem}. 

\subsubsection{Two kinds of critical points and their spectral curves \label{SubSubSectionTwoCriticalPoints}}

The critical points of the multi-cut two-matrix models are classified by these lax operators $(\bP,\bQ)$. 
Fractional-superstring $(\hat p,\hat q)$ critical points are originally discussed in \cite{irie2} 
as the critical points which realize the operator contents of $(\hat p,\hat q)$ minimal fractional superstring theory. 
From the quantitative analysis of these critical points \cite{CISY1} 
(which is the multi-cut extension of \cite{DKK}), it was found that there are two kinds of 
fractional-superstring critical points which are called 
{\em $\omega^{1/2}$-rotated models} and {\em real-potential models}. 
The parameter $\epsilon$ in Eq.~\eq{ScalingArgument} 
then is related with these two choices of the fractional-superstring critical points: 
\begin{align}
\epsilon = 
\left\{
\begin{array}{cl}
0 & \text{: $\omega^{1/2}$-rotated models} \cr
1 & \text{: real-potential models}
\end{array}
\right..
\end{align}
These two kinds of solutions are perturbatively distinct when $k\in 4\mathbb Z$; non-perturbatively always distinct. Here we summarize the results of 
critical-potential analysis \cite{CISY1} and the weak-coupling spectral curves \cite{CIY1}: 

\begin{itemize}
\item[1.]
The first kind is derived from {\em $\omega^{1/2}$-rotated-potential models} \cite{CISY1}, 
and given as
\begin{align}
\bP(t;\del) = \Gamma\, \del^{\hat p} + \sum_{n=1}^{\hat p} H_n^{(\rm F_kP)} (t)\,\del^{\hat p-n},\qquad 
\bQ(t;\del) = \Gamma\, \del^{\hat q} + \sum_{n=1}^{\hat q} H_n^{(\rm F_kQ)} (t)\,\del^{\hat q-n},
\end{align}
with the shift matrix $\Gamma$, 
\begin{align}
\Gamma= 
\begin{pmatrix}
0 & 1 \cr 
 & 0 & 1 \cr
 & & \ddots & \ddots \cr
 & & &0& 1 \cr
1& & & & 0
\end{pmatrix},
\end{align}
and the real functions $H_n^{(\rm F_kP)} (t)$ and $H_n^{(\rm F_kQ)} (t)$. 
The weak coupling analysis $g_{\rm str}\to 0$ results in the following simultaneous eigenvalues of the Lax operators 
$\bP(t;\del) \simeq \diag_{j=1}^k\bigl(P^{(j)}(t;z)\bigr),
\bQ(t;\del) \simeq \diag_{j=1}^k\bigl(Q^{(j)}(t;z)\bigr)$ \cite{CIY1}:
\begin{align}
&\underline{\text{Cosh solution:}}\nn\\
&\qquad P^{(j)} =\lambda^{\hat p} \cosh\bigl(\hat p \tau + 2\pi i \frac{j-1}{k}\bigr),
\qquad Q^{(j)} = \lambda^{\hat q} \cosh\bigl(\hat q \tau + 2\pi i \frac{j-1}{k}\bigr); \\
&\underline{\text{Sinh solution:}}\nn\\
&\qquad P^{(j)} = \lambda^{\hat p} \sinh\bigl(\hat p \tau + 2\pi i \frac{j-1}{k}\bigr),\qquad 
Q^{(j)} = \lambda^{\hat q} \sinh\bigl(\hat q \tau + 2\pi i \frac{j-1}{k}\bigr),
\end{align}
with 
\begin{align}
\lambda \equiv t^{\frac{1}{\hat p+\hat q-1}},\qquad 
z \equiv \lambda^{-(\hat p+\hat q)} g_{str}\, t\, \del_t = \cosh(\tau).  \label{DefinitionLambdaAndZ}
\end{align}
Their algebraic equations are expressed as%
\footnote{$T_p(\cosh(\tau))\equiv \cosh(p\tau)$ stands for the Chebyshev polynomial of the first kind}
\begin{align}
&\underline{\text{Cosh solution:}}\qquad F(P,Q)= T_p(Q/\lambda^{\hat q})
- T_q(P/\lambda^{\hat p})=0; \\
&\underline{\text{Sinh solution:}}\qquad F(P,Q)=T_p(-i Q/\lambda^{\hat q}) 
- (-1)^{\frac{q-p}{2}} T_q(-i P/\lambda^{\hat p}) =0.
\end{align}
Here the indices $(p,q)$ are identified with 
the labeling of minimal fractional superconformal models \cite{irie2} 
and defined by 
\begin{align}
(p,q) = (\hat k \hat p,\hat k \hat q), \label{PQofFSCFT1}
\end{align}  
with introducing the following integers $d_{\hat q-\hat p}, \,\hat k$ and $\eta$, 
\begin{align}
d_{\hat q-\hat p} \equiv \text{g.c.d.}\bigl(\hat q-\hat p, k\bigr),\qquad 
k = \hat k \times d_{\hat q-\hat p}, \qquad \hat q-\hat p = \eta \times d_{\hat q-\hat p}.
\label{PQofFSCFT2}
\end{align}
\item[2.]
The other kind is derived from {\em real-potential models} \cite{CISY1} and given by replacing 
the shift matrix $\Gamma$ by the twisted shift matrix $\Gamma_{(\rm real)}$ as follows:
\begin{align}
\bP(t;\del) = \Gamma_{(\rm real)}\, \del^{\hat p} + \sum_{n=1}^{\hat p} H_n^{(\rm R_kP)} (t)\,\del^{\hat p-n},\qquad 
\bQ(t;\del) = \Gamma_{(\rm real)}\, \del^{\hat q} + \sum_{n=1}^{\hat q} H_n^{(\rm R_kQ)} (t)\,\del^{\hat q-n},
\end{align}
with the twisted shift matrix $\Gamma_{(\rm real)}$, 
\begin{align}
\Gamma_{(\rm real)}= 
\begin{pmatrix}
0 & 1 \cr 
 & 0 & 1 \cr
 & & \ddots & \ddots \cr
 & & &0& 1 \cr
-1& & & & 0
\end{pmatrix},
\end{align}
and the real functions $H_n^{(\rm R_kP)} (t)$ and $H_n^{(\rm R_kQ)} (t)$. 
The weak coupling analysis $g_{\rm str}\to 0$ results in the following eigenvalues of the Lax operators 
$\bP(t;\del) \simeq \diag_{j=1}^k\bigl(P^{(j)}(t;z)\bigr)$, 
$\bQ(t;\del) \simeq \diag_{j=1}^k\bigl(Q^{(j)}(t;z)\bigr)$ \cite{CIY1}:
\begin{align}
&\underline{\text{Cosh solution:}}\nn\\
&\qquad P^{(j)} = \lambda^{\hat p} \cosh\bigl(\hat p \tau + 2\pi i \frac{2j-1}{2k}\bigr),
\qquad Q^{(j)} = \lambda^{\hat q} \cosh\bigl(\hat q \tau + 2\pi i \frac{2j-1}{2k}\bigr); \\
&\underline{\text{Sinh solution:}}\nn\\
&\qquad P^{(j)} = \lambda^{\hat p} \sinh\bigl(\hat p \tau + 2\pi i \frac{2j-1}{2k}\bigr),\qquad 
Q^{(j)} = \lambda^{\hat q} \sinh\bigl(\hat q \tau + 2\pi i \frac{2j-1}{2k}\bigr),
\end{align}
with the parameters in Eqs.~\eq{DefinitionLambdaAndZ}. 
Their algebraic equations are expressed as 
\begin{align}
&\underline{\text{Cosh solution:}}\qquad F(P,Q)= T_p(Q/\lambda^{\hat q})
-(-1)^{\{ \eta \}}\,T_q(P/\lambda^{\hat p})=0; \\
&\underline{\text{Sinh solution:}}\qquad F(P,Q)=T_p(-i Q/\lambda^{\hat q}) 
- (-1)^{\{\eta\}+\frac{q-p}{2}} T_q(-i P/\lambda^{\hat p}) =0.
\end{align}
Here $\eta$ is defined in Eq.~\eq{PQofFSCFT2}. 
\end{itemize}

\subsubsection{Superposition of perturbative string theories}

A nontrivial feature of these explicit expressions \cite{CIY1} is that these algebraic equations $F(P,Q)=0$ are 
generally factorized into several irreducible pieces of the algebraic curves $F^{(j)}(P,Q)=0,\, (j=1,2,\cdots,k)$:%
\footnote{Each equation $F^{(j)}(P,Q)=0$ corresponds to an eigenvalue of the Lax pair $(\bP,\bQ)$. 
Then the upper index of the equations satisfies $F^{(k-j)}(P,Q)=F^{(j)}(P,Q)=F^{(j + k)}(P,Q)=0$. }
\begin{align}
F(P,Q) = \prod_{j=1}^{\lfloor \frac{k}{2}\rfloor+1} F^{(j)}(P,Q)=0,
\end{align}
and these irreducible pieces are equivalent to 
the algebraic curves of the fractional-superstring critical points (of the matrix models) 
with the smaller number of cuts. 
For instance, some are equivalent to the curves of minimal bosonic string theory and some are
those of minimal type 0 superstring theory.

The authors of \cite{CIY1} derived the algebraic equation as the following limit of the Baker-Akhiezer system: 
\begin{align}
F(P,Q) = 0\qquad \text{s.t.}\qquad 
F\Bigl(\zeta, g_{\rm str}\frac{\del}{\del \zeta} \Bigr)\Psi(t;\zeta) = 0\qquad g_{\rm str}\to 0. 
\end{align}
It was shown by the general arguments of topological recursions \cite{EynardLoop} 
that the algebraic equation has enough information to recover the perturbative corrections of the string theories. 
This means that we can reconstruct ``perturbative Baker-Akhiezer functions'' from the perturbative 
pieces of the algebraic equation: 
\begin{align}
\Psi_{\rm pert}^{(j)}(t;\zeta):\qquad 
F^{(j)}\Bigl(\zeta, g_{\rm str}\frac{\del}{\del \zeta} \Bigr)\Psi_{\rm pert}^{(j)}(t;\zeta) = 0\qquad g_{\rm str}\to 0, 
\end{align}
the leading behavior of which is given by one of the branches of the algebraic equation $\mathcal R^{(j)}(t;\zeta)$, 
$F\bigl(\zeta,\mathcal R^{(j)}(t;\zeta)\bigr)=0$: 
\begin{align}
\Psi_{\rm pert}^{(j)}(t;\zeta) \asymeq v^{(j)}(t;\zeta) \, \exp\Bigl[
{g_{\rm str}^{-1}\int^\zeta d\zeta' \mathcal R^{(j)}(t;\zeta')}
\Bigr] + \cdots, \quad g_{\rm str}\to 0,
\label{BehaviorOfPerturbativeBakerAkhiezerFunctions}
\end{align}
where $v^{(j)}(t;\zeta)$ is a proper $k$-order vector-valued function.%
\footnote{
These perturbative Baker-Akhiezer functions also receive non-perturbative corrections by ZZ-branes of the 
corresponding perturbative string theory. The general form of these functions is universal 
and is expressed by theta functions on the spectral curve \cite{EynardMarino} 
(in other words, they are the $\tau$-functions of the corresponding integrable hierarchy). }
Interestingly, these wave functions $\Psi_{\rm orth}^{(j)} \, (j=1,2,\cdots,k)$ 
are completely decouple to each other in all-order perturbation theory, 
since the algebraic equations are factorized into these irreducible pieces. 
Clearly, the exact Baker-Akhiezer function $\Psi(t;\zeta)$ and the perturbative Baker-Akhiezer functions $\Psi_{\rm pert}^{(j)}(t;\zeta)$ are different, 
but an asymptotic expansion of the exact Baker-Akhiezer function $\Psi(t;\zeta)$ in $g_{\rm str}$ 
is given by a superposition of 
the perturbative Baker-Akhiezer functions $\Psi_{\rm pert}^{(j)}(t;\zeta)$: 
\begin{align}
\Psi(t;\zeta) \asymeq \sum_{j} c_j \, \Psi_{\rm pert}^{(j)}(t;\zeta), 
\qquad g_{\rm str}\to 0,\quad \text{with} \quad(t,\zeta) \in {}^\exists D \subset \mathbb C^2. \label{PsiByAsymeqOfPerBAFunc}
\end{align}
This result suggests that the total wave function of the fractional superstring theory 
is a superposition of various perturbative minimal string theories \cite{CIY1}! 
One of our motivations in this paper 
is to further investigate this intriguing point from the non-perturbative 
point of views. 

In particular, from the perturbation theory, 
there is no way to fix the relative coefficients $\{c_j\}_{j=1}^k$. 
This is the orthonormal-polynomial version of the non-perturbative ambiguity, and 
it is clear that these coefficients $\{c_j\}_{j=1}^k$ cannot take arbitrary values, 
even though these parameters $\{c_j\}_{j=1}^k$ are an analogy of the theta parameters in QCD. 
The correct values are obtained in the $\mathbb Z_k$-symmetric critical points \cite{CIY2} 
and we will show the fractional-superstring cases in Section \ref{SubsectionExplicitSolutions}. 
Interestingly, as is shown in \cite{CIY2} and as we will see in Section \ref{SubsectionExplicitSolutions}, 
determination of these coefficients $\{c_j\}_{j=1}^k$ are equivalent to fixing the D-instanton fugacities. 

\subsubsection{The multi-cut geometry and Stokes phenomena \label{SubsubsectionMCgeometryAndStokesPhen}}

By an analytic continuation of $\zeta$, 
the most dominant perturbative Baker-Akhiezer functions in the asymptotic expansion 
Eq.~\eq{PsiByAsymeqOfPerBAFunc} will change in a discontinuous way. 
Such discontinuities generally appear when one crosses the Stokes lines: 
\begin{align}
{\rm SL}_{j,l} \equiv 
\Bigl\{\zeta \in \mathbb C\, ; \,{\rm Re} \int^\zeta d\zeta' \Bigl[\mathcal R^{(j)}(t;\zeta')- \mathcal R^{(l)}(t;\zeta')\Bigr] = 0 \Bigr\},\quad (j,l=1,2,\cdots,k).  \label{GeneralStokesLines}
\end{align}
In particular, the Stokes lines can be interpreted as {\em physical cuts} of the eigenvalue distribution function $\rho(\lambda)$ of Eq.~\eq{DefinitionOfResolventAndEigenvalueDensityFunction},%
\footnote{The physical cuts can be curved lines if the matrix-model potential includes complex coefficients. 
However the eigenvalue distribution function along the lines should be a real function. This requirement 
is guaranteed by the definition of the Stokes lines, Eq.~\eq{GeneralStokesLines} (See \cite{CIY2}). }
if the exact Baker-Akhiezer 
function is given by the scaling orthonormal polynomials $\Psi_{\rm orth}(t;\zeta)$ of 
Eq.~\eq{ScalingOrthonormalPolynomials} (and Eq.~\eq{ResolventOrthonormalPoly}), 
\begin{align}
&\Psi(t;\zeta) = \Psi_{\rm orth}(t;\zeta) \asymeq v(t;\zeta) \exp\Bigl[g_{\rm str}^{-1}\int^\zeta d\zeta' \,\mathcal R(\zeta')\Bigr]+\cdots, \nn\\
&\qquad \qquad \qquad \qquad g_{\rm str}\to 0\qquad \text{with} \qquad(t,\zeta) \in {}^\exists D \subset \mathbb C^2. 
\end{align}
Here $\mathcal R(\zeta)$ is the scaling resolvent operator, 
$N dx R(x) = g_{\rm str}^{-1}\, d\zeta\, \mathcal R(\zeta)$. 
From the viewpoint of the large $N$ resolvent, Eq.~\eq{DefinitionOfResolventAndEigenvalueDensityFunction}, 
it is clear that the physical cuts (around $\zeta\to\infty$) 
appear along the particular angles, $\zeta \to \infty \times e^{i\chi_n}$:%
\footnote{Note that there is a nontrivial rotation of angle by $\epsilon$, $x \sim \omega^{\epsilon/2}\zeta$, 
in Eq.~\eq{ScalingArgument}. }
\begin{align}
\chi_n=\chi_0 + \frac{2\pi n}{k}, \qquad \chi_0 = \frac{\pi}{k}(1-\epsilon),\qquad 
\bigl(n=0,1,2,\cdots,k-1\bigr). \label{DefinitionOfChiAngleInMCBC}
\end{align}
However this consideration is not straightforward from the Baker-Akhiezer function system Eq.~\eq{DESystem}
and implies non-trivial physical constraints on the coefficients $\{c_j\}_{j=1}^k$ 
in Eq.~\eq{PsiByAsymeqOfPerBAFunc} 
and also on the Stokes phenomena which relate the coefficients in different regions $D$ 
of the asymptotic expansions 
Eq.~\eq{PsiByAsymeqOfPerBAFunc}. This is the multi-cut boundary condition proposed in \cite{CIY2}.

This consideration in the fractional-superstring critical points is quite interesting because 
the perturbative minimal string theories described by the irreducible algebraic eqations, 
$F^{(j)}(P,Q)=0,\, (j=1,2,\cdots,k)$ are completely decouple in the all-order perturbation theory. 
Therefore, the discontinuous jumps of the wave function $\Psi_{\rm orth}(t;\zeta)$ are the jumps 
between distinct perturbative string theories (i.e.~not the jumps 
within the same perturbative string theory). Therefore, this means that 
different perturbative string theories are distributed in different regimes of $\zeta$ 
and they are connected by Stokes phenomena! This idea was first introduced in \cite{CIY1} 
in order to add physical cuts which disappear in the spectral curves $F(P,Q)=0$. 
A quantitative demonstration of this idea was then carried out 
in the $\mathbb Z_k$ symmetric critical points \cite{CIY2}. 
Here we now analyze what actually happens in the fractional-superstring critical points.

\subsection{Stokes phenomena in the multi-cut critical points}

The general frameworks of the Stokes phenomena in the multi-cut critical points 
are developed in Section 3 of \cite{CIY2} (a review of basics of Stokes phenomena 
with relevant references is in \cite{ItsBook} or in Section 2 of \cite{CIY2}). 
Here we will not repeat the discussions but 
the relevant results are briefly summarized in order to fix the notations and conventions. 
Some new features appearing in the fractional-superstring critical points are also mentioned. 
For sake of simplicity, the discussion is restricted to $\hat p=1$.

In the case of $\hat p=1$, the Lax operator $\bP(t;\zeta)$ is generally given as 
\begin{align}
\bP(t;\del) = A\, \del + H(t),\qquad \det A\neq 0, 
\end{align}
and therefore the differential equation system Eqs.~\eq{DESystem} is rewritten as 
the following ODE system of the Zakharov-Shabat eigenvalue problem \cite{ZS,AKNS}: 
\begin{align}
g_{\rm str}\frac{\del}{\del t }\Psi(t;\zeta)&= \mathcal P(t;\zeta)\, \Psi(t;\zeta) \equiv A^{-1}\Bigl(\zeta - H(t)\Bigr)\,\Psi(t;\zeta); \\
g_{\rm str} \frac{\del}{\del \zeta}\Psi(t;\zeta) &= \mathcal Q(t;\zeta) \, \Psi(t;\zeta)
\equiv \bQ(t;\del)\, \Psi(t;\zeta). \label{FSSTODEsystem}
\end{align}
The Douglas (i.e.~string) equation \cite{DouglasGeneralizedKdV} is 
now given by the following Zakharov-Shabat zero-curvature form of $(\mathcal P,\mathcal Q)$: 
\begin{align}
\bigl[\bP(t;\del),\bQ(t;\del)\bigr]=g_{\rm str} I_k\quad \Leftrightarrow \quad 
\bigl[g_{\rm str}\frac{\del}{\del t}-\mathcal P(t;\zeta),g_{\rm str}\frac{\del}{\del \zeta}-\mathcal Q(t;\zeta)\bigr]=0. 
\end{align}
This system is referred to as $k\times k$ isomonodromy system, 
since this differential-equation system preserves the monodromy data and therefore Stokes data 
of Eq.~\eq{FSSTODEsystem} 
with respect to the flows of $(t;\zeta)$.

The leading behavior of the ordinary differential equation, Eq.~\eq{FSSTODEsystem}, is given by 
\begin{align}
g_{\rm str}\frac{\del \Psi(t;\zeta)}{\del \zeta} = \bigl(A^{-\gamma} \zeta^{r-1} + \cdots\bigl)\Psi(t;\zeta),\qquad \gamma=r-2. 
\end{align}
Here $r=\hat q+1$ is the Poincar\'e index of the essential singularity at $\zeta \to \infty$. 
We consider two choices of the matrix $A$:%
\footnote{In this paper, ``$\simeq$'' indicates equality up to some similarity transformation. }
\begin{align}
A=
\left\{
\begin{array}{ll}
\Gamma \simeq \Omega & \text{: $\omega^{1/2}$-rotated models} \cr
\Gamma_{\rm (real)} \simeq \omega^{1/2}\Omega & \text{: real-potential models}
\end{array}
\right.,\qquad \Omega=\diag_{j=1}^k\bigl(\omega^{j-1}\bigr). 
\end{align}
As in \cite{CIY2}, we also restrict ourselves to the cases of ${\rm g.c.d.}\, (k,\gamma)=1$ 
in order to avoid complexity 
due to degeneracy of the exponents and appearance of the subdominant exponents. 
Interestingly, this condition coincides with the following condition on $d_{\hat q-\hat p}$ in \eq{PQofFSCFT2}:
\begin{align}
d_{\hat q-\hat p}={\rm g.c.d.}\, (k,\hat q-1) =1.
\end{align}
From the fractional-superstring viewpoint, this condition means the restriction to 
the critical points which are free from the $\mathbb Z_k \times \mathbb Z_k$-orbifolding \cite{irie2}. 
This kind of orbifoldings was first introduced in the two-cut critical points of the two-matrix models \cite{fi1} 
(i.e.~$\mathbb Z_2$-orbifolding in the cases) in order to keep the correspondence 
with perturbative type 0 minimal superstring theories 
of odd sequence $(\hat q,\hat p \in 2\mathbb Z+1)$. 
Therefore, the above coincidence suggests that {\em 
this $\mathbb Z_k \times \mathbb Z_k$-orbifolding procedure 
is naturally encoded in the non-perturbative Stokes-phenomenon structure  
and can be imposed by just completely identifying the degenerate exponents.} 
This is very interesting observation but the investigation along this direction is out of scope 
in this paper. 

The basic ingredients of Stokes phenomena are following: 

\subsubsection{Asymptotic expansions}

This ODE system Eq.~\eq{FSSTODEsystem} has the $k$ independent order-$k$ column vector solutions 
$\Psi^{(j)}(t;\zeta)$, ($j=1,2,\cdots,k$), and they are often expressed as the following matrix: 
\begin{align}
\Psi(t;\zeta) \equiv \Bigl(\Psi^{(1)}(t;\zeta),\cdots, \Psi^{(k)}(t;\zeta)\Bigr). \label{MatrixNotationOfODESolutions}
\end{align}
The Stokes phenomena discussed in this paper originates from 
the asymptotic expansion around $\zeta\to \infty$. In term of the matrix notation Eq.~\eq{MatrixNotationOfODESolutions}, with a proper change of normalization of the solutions, 
the expansion can always be expressed as 
\begin{align}
\Psi_{asym}(t;\zeta) 
&\equiv Y(t;\zeta) \, e^{\frac{1}{g_{\rm str}}\varphi(t;\zeta)} \nn\\
&\equiv  \Bigl[I_k +\sum_{n=1} ^\infty \frac{Y_n(t)}{\zeta^n}\Bigr]\times 
\exp \Bigl[\frac{1}{g_{\rm str}}
\Bigl(\varphi_0 \ln \zeta - \sum_{m=-r,\neq 0}^\infty \frac{\varphi_m(t)}{m\,\zeta^m} \Bigr)\Bigr]. 
\label{AsymptoticExpansionOfPsiInMatrixBasis}
\end{align}
The leading exponents are written as 
\begin{align}
\varphi(t;\zeta) = \frac{A^{-\gamma}}{r} \zeta^r + O(\zeta^{r-1}),\qquad \zeta\to \infty. 
\end{align}
From the hermiticity of the matrix models, these matrix solutions are always real functions \cite{CISY1}. 
This basis of the solutions is called {\em the matrix-model basis} (or {\em the $\Gamma$-basis}). 
On the other hand, the Stokes phenomena are always discussed in {\em the diagonal basis}
(or {\em the $\Omega$-basis}): 
\begin{align}
\widetilde \Psi_{asym}(t;\zeta) &\equiv
\widetilde Y(t;\zeta) \, e^{\frac{1}{g_{\rm str}}\widetilde \varphi(t;\zeta)} \nn\\
&\equiv  \Bigl[I_k +\sum_{n=1} ^\infty \frac{\widetilde Y_n(t)}{\zeta^n}\Bigr]\times 
\exp \Bigl[\frac{1}{g_{\rm str}}\Bigl(\widetilde \varphi_0 \ln \zeta - \sum_{m=-r,\neq 0}^\infty \frac{\widetilde \varphi_m(t)}{m\,\zeta^m} \Bigr)\Bigr] \nn\\
&\equiv U^\dagger \, \Psi_{asym}(t;\zeta) \, U, 
\label{AsymptoticExpansionOfPsiInDiagonalBasis}
\end{align}
where the unitary matrix $U$ is given as 
\begin{align}
U_{jl}=\frac{1}{\sqrt{k}} \omega^{(j-1+\epsilon/2)(l-1+\epsilon/2)},\qquad A\, U = U\, \bigl(\omega^{\epsilon/2}\Omega\bigr).  \label{UnitaryMatrixOfMatrixDiagonalBasis}
\end{align}
Therefore, the exponents are diagonal matrices: 
$\widetilde \varphi(t;\zeta) = \diag_{j=1}^k \bigl(\varphi^{(j)}(t;\zeta)\bigr)$, 
and the leading behavior is always the same as the exponents of the perturbative 
Baker-Akhiezer functions Eq.~\eq{BehaviorOfPerturbativeBakerAkhiezerFunctions},
\begin{align}
\varphi^{(j)}(t;\zeta)- \int^\zeta d\zeta' \, \mathcal R^{(j)}(t;\zeta') = O(\zeta^0), 
\qquad \zeta \to \infty. 
\end{align}

\subsubsection{Stokes sectors}

The reconstruction of the exact solutions from these asymptotic expansions is 
achieved in specific angular domains called {\em Stokes sectors} $D_n$.
This can be defined by {\em Stokes lines} Eq.~\eq{GeneralStokesLines}, 
especially their leading behavior of $\zeta \to\infty$: 
\begin{align}
{\rm sl}_{j,l} &\equiv \Bigl\{ \zeta \in \mathbb C\,;\, {\rm Re}\bigl[\bigl(\varphi_{-r}^{(j)}-\varphi_{-r}^{(l)}\bigr)\zeta^r\bigr]=0\Bigr\} \nn\\
&=\Bigl\{ \zeta \in e^{i \theta_{j,l}^{(n)}} \, \mathbb R \,;\, 
\theta_{j,l}^{(n)} = \frac{k n + \gamma (j+l-2+\epsilon)}{rk}\pi, \, n\in \mathbb Z\Bigr\}, \label{StokesLineFormula}
\end{align}
where $\varphi^{(j)}_{-r}$ is the leading exponent of $\varphi^{(j)}(t;\zeta) = (1/r)\,\varphi^{(j)}_{-r}\zeta^r + \cdots$, and $\gamma=r-2$. 
Because of ${\rm g.c.d.}\, (k,\gamma)=1$, the minimal angular domains in between 
the Stokes lines, $\delta D_n$, are given by%
\footnote{In this paper, we use the following notation of angular domain: $D(a,b) \equiv \{a<\arg (\zeta) < b\}$. }
\begin{align}
\delta D_n = D\bigl(\frac{\pi (n-1)}{kr}, \frac{\pi n}{kr}\bigr), \label{DefinitionOfSegments}
\end{align}
which are referred to as {\em segments}. Therefore, the Stokes sectors are defined as 
the angular domains which consist of $(k+1)$ different segments: 
\begin{align}
D_n = D\bigl(\frac{\pi (n-1)}{kr}, \frac{\pi (n+k)}{kr}\bigr),
\end{align}
which are also called {\em fine Stokes sectors}.%
\footnote{This choice of Stokes sectors is most fundamental. For other descendant Stokes sectors, see \cite{CIY2}. } In each Stokes sector, one can uniquely reconstruct the exact (canonical) solutions 
$\widetilde \Psi_n(t;\zeta)$ from the asymptotic expansion Eq.~\eq{AsymptoticExpansionOfPsiInDiagonalBasis}: 
\begin{align}
\widetilde \Psi_n(t;\zeta) \asymeq \widetilde \Psi_{\rm asym}(t;\zeta), \qquad \zeta \to \infty \in D_n. 
\end{align}
However, in general, these exact solutions have different normalizations: 
\begin{align}
\widetilde \Psi_{n+1}(t;\zeta) = \widetilde \Psi_n(t;\zeta) S_n. 
\end{align}
This behavior of asymptotic expansion is called {\em Stokes phenomena} in this ODE system 
and the matrix $S_n$ is called {\em (fine) Stokes matrix}. 
The total number of segments is $2rk$ and is equal to that of fine Stokes sectors. 

\subsubsection{Profiles and Stokes multipliers}

Reading the positions of non-trivial components in the Stokes matrices, $S_n = (s_{n,i,j})$, 
is always a tedious problem, although there is a basic theorem which can be applied to every ODE system in principle.%
\footnote{See Theorem 2 in \cite{CIY2}, for example. }
Here, however, we use more efficient way which was found in \cite{CIY2} with {\em profile of dominant exponents}. 
The profile of dominant exponents $\mathcal J$ for $(k,r;\gamma)$ is 
a sequence of integer numbers $\{j_{l,n}\}$ of the following type:
\begin{align}
\mathcal J \equiv 
\left[
\begin{array}{c}
J_{2rk-1} \cr
\hline
\vdots  \cr
\hline
J_1 \cr
\hline
J_0
\end{array}
\right] 
\equiv \left[
\begin{array}{c|c|c|c}
j_{2rk-1,1} & j_{2rk-1,2} & \cdots & j_{2rk-1,k} \cr
\hline
\vdots & \vdots & & \vdots \cr
\hline
j_{1,1} & j_{1,2} & \cdots & j_{1,k} \cr
\hline
j_{0,1} & j_{0,2} & \cdots & j_{0,k}
\end{array}
\right]. 
\end{align}
Each row vector $J_l$ corresponds to a segment $\delta D_l$ of Eq.~\eq{DefinitionOfSegments}, 
and the numbers in $J_l = 
\left[
\begin{array}{c|c|c|c}
j_{l,1} &j_{l,2} & \cdots & j_{l,k}
\end{array}
\right]$ 
describes the profile of dominant exponents in the segment $\delta D_l$:
\begin{align}
{\rm Re}\bigl[
\varphi_{-r}^{(j_{l,1})} \zeta^r\bigr]<
{\rm Re}\bigl[\varphi_{-r}^{(j_{l,2})} \zeta^r \bigr]<\cdots<
{\rm Re}\bigl[
\varphi_{-r}^{(j_{l,k})} \zeta^r\bigr],
\qquad \zeta \in \delta D_l.
\end{align}
The profile is expressed as $\mathcal J_{(k,r)}^{(\epsilon)}$ in order to explicitly show the indices $(k,r)$ 
and $\epsilon = 0,1$. In this paper, $\gamma$ is always $\gamma = r-2$. 
As an example, a part of the profile $\mathcal J_{(7,3)}^{(0)}$ ($7$-cut $(\hat p,\hat q)=(1,2)$ $\omega^{1/2}$-rotated fractional-superstring critical point) is shown:
\begin{align}
\mathcal J_{(7,3)}^{(0)} = \left[
\begin{array}{c|c|c|c|c|c|c}
\vdots & \vdots & \vdots & \vdots & \vdots & \vdots & \vdots \cr
\hline
7 & (6 & 1) & (5 & 2) & (4 & 3) \cr
\hline
(6 & 7) & (5 & 1) & (4 & 2) & 3 \cr
\hline
6 & (5 & 7) & (4 & 1) & (3 & 2) \cr
\hline
(5 & 6) & (4 & 7) & (3 & 1) & 2 \cr
\hline
5 & (4 & 6) & (3 & 7) & (2 & 1) \cr
\hline
(4 & 5) & (3 & 6) & (2 & 7) & 1 
\end{array}
\right]. \label{ProfileIn7CutR3Omega}
\end{align}
Here we put the parentheses as $(i|j)$ in the profile to indicate the pairs $(i,j)$ which 
change the dominance in the next segment (i.e.~there is a Stokes line of $sl_{i,j}$ just above the segment). 
We also use the notation $(i|j)_l$ to show the segment $J_l$ which the pair belongs to: $(i|j)_l \in J_l$. 
Then the non-trivial claim (Theorem 4 in \cite{CIY2}) is 
\begin{align}
\text{$s_{l,i,j}$ is non-trivial component of $S_l$} \qquad \Leftrightarrow \qquad (j|i)_l \in J_l. 
\end{align}
The others off-diagonal components are zero, $s_{l,i,j}=0$. From the standard theorem, 
the diagonal components are always $s_{l,i,i}=1\ (i=1,2,\cdots,k)$. 

\subsubsection{Formula of profile components}

A simple formula for the integer numbers in the profile $j_{l,n}$ is also obtained in \cite{CIY2} 
(Theorem 3) for the $\omega^{1/2}$-rotated models 
and the formula for the real-potential models is also obtained by a minor change:%
\footnote{Note that the first exponent $\varphi_{-r}^{(1)} \zeta^r = \omega^{-\gamma/2} \zeta^r$ 
becomes the largest when $\arg(\zeta) = \dfrac{\pi}{kr} \gamma$. Also note that 
the effect of $\epsilon$ is just a total shift of $\theta_{j,l}^{(n)}$ in Eq.~\eq{StokesLineFormula}. }
\begin{align}
j_{l,n}^{(\epsilon)} \equiv 1 + \biggl(\Bigl\lfloor \frac{(l-\epsilon \gamma)}{2}\Bigr\rfloor 
+ (-1)^{k+(l-\epsilon\gamma)+n} \Bigl\lfloor \frac{k-n+1}{2}\Bigr\rfloor\biggr) m_1,\quad 
\mod k, \label{FormulaForProfile}
\end{align}
where $m_1$ is obtained by the Euclidean algorithm 
of $k n_1 + \gamma m_1 = 1$. Interestingly, the difference of the formula for the Stokes phenomena 
between the $\omega^{1/2}$-rotated models and the real-potential models is just a shift of the angular 
coordinate and essentially they both have the same structure, 
even though these models are distinct from the viewpoint of perturbation theory. 
This implies that different perturbative non-critical string theories are simply unified in the strong-coupling non-critical M-theory description, 
as is in the critical 11-dimensional M theory. 

\subsubsection{Three basic constraints}

There are basic three (or two) constraints on the Stokes multipliers. 
The original three constraints in the two-cut cases 
can be found in \cite{ItsBook},%
\footnote{See also Section 2 in \cite{CIY2} since the convention used in the matrix models 
is different. }
and also the extension to the multi-cut cases is in \cite{CIY2} 
with referring to the matrix-model results \cite{CISY1}. 
A major difference in the fractional-superstring critical points is that there is no $\mathbb Z_k$ symmetry 
(or there is only $\mathbb Z_2$ symmetry when $k\in 2\mathbb Z$). 
These constraints are also obtained in the similar way as in \cite{CIY2}:

\paragraph{$\mathbb Z_2$-symmetry condition (only when $k\in 2\mathbb Z$): }
\begin{align}
S_{n+rk} = \Gamma^{-\frac{k}{2}} S_n \Gamma^{\frac{k}{2}},\qquad \bigl(n=0,1,\cdots, 2rk-1\bigr). 
\label{Z2SymmetryCondition}
\end{align}

\paragraph{The hermiticity conditions: }
\begin{align}
S_n^*= \Delta\Gamma^{(1-\epsilon)}\, S_{(2r-1)k-n}^{-1}\, \Gamma^{-(1-\epsilon)}\Delta, 
\qquad \bigl(n=0,1,\cdots, 2rk-1\bigr), 
\label{HermiticityCondStokes}
\end{align}
with the matrix $\Delta$, $\Delta_{i,j}=\delta_{i,k-j+1}$. Note that the unitary matrix $U$ 
of Eq.~\eq{UnitaryMatrixOfMatrixDiagonalBasis} satisfies $U^2 = (-1)^\epsilon\Delta\Gamma^{(1-\epsilon)}$. 
In terms of the elements, they are expressed as 
\begin{align}
s_{n,i,j}^* = - s_{-k-n, 2-\epsilon -i,2-\epsilon -j},\qquad  (j|i)_n \in J_n. 
\end{align}

\paragraph{The monodromy free condition: }
\begin{align}
S_0 \cdot S_1 \cdots S_{2rk-1} 
= I_k. \label{OriginalMonodromyFreeCondition}
\end{align}
In the case of $k\in 2\mathbb Z$, this is also expressed as 
\begin{align}
\Bigl[(S_0\cdots S_{rk-1})\Gamma^{-\frac{k}{2}} \Bigr]^2 = I_k. 
\end{align}
In the case of $k\in 2\mathbb Z+1$, this is also expressed as 
\begin{align}
\mathcal S^* = \Delta \Gamma^{(1-\epsilon)} \, \mathcal S \, \Delta \Gamma^{(1-\epsilon)}, 
\end{align}
with $\mathcal S \equiv  S_{-\frac{k+1}{2}+1}\,  S_{-\frac{k+1}{2}+2}\, \cdots S_0 \, S_1\cdots S_{rk-\frac{k+1}{2}}$. 

\section{The multi-cut boundary conditions (BC) \label{SectionMCBC}}

In this section, we evaluate the multi-cut boundary conditions 
in the fractional-superstring critical points. 
The multi-cut boundary condition is first proposed in \cite{CIY2} (Definition 10) 
and also mentioned around Eq.~\eq{DefinitionOfChiAngleInMCBC} in this paper. 

One of the main differences from the $\mathbb Z_k$ symmetric cases \cite{CIY2} is 
that there are a huge number of Stokes multipliers.%
\footnote{For example, there are seven times more than the multipliers shown in the profile, Eq.~\eq{ProfileIn7CutR3Omega}, in the 7-cut $r=3$ case. } It is because 
the smallest Poincar\'e index $r$ is $r=3$ (there is no $(\hat p,\hat q)=(1,1)$ critical point) 
and there is no $\mathbb Z_k$ symmetry. 
This causes a predicament in direct analysis of the algebraic equations of Stokes multipliers. 
Therefore, we should invent some systematic way to analyze these large systems. 

An important property of the formula for the profile components Eq.~\eq{FormulaForProfile} is 
the following vertical shift in the profile: 
\begin{align}
j_{l + 2a, n}^{(\epsilon)} = j_{l, n}^{(\epsilon)} + a\,m_1.  \label{ProfileShiftRule}
\end{align}
This implies the following fact: 
\begin{align}
&\text{$s_{l,i,j}$ is non-trivial (i.e.~${}^\exists (j|i)_l\in J_l$)} \nn\\ 
&\Leftrightarrow\quad 
\text{$s_{l+2a,i+am_1,j+am_1}$ is non-trivial (i.e.~${}^\exists (j+am_1|i+am_1)_{l+2a}\in J_{l+2a}$)}. 
\end{align}
Therefore, once we know the existence of non-trivial Stokes multipliers of $s_{l,i,j}$ ($l=0,1,2,\cdots,2r-1$), then 
one can guarantee the existence of the other Stokes multipliers by using the above shift rule. 
This rule is referred to as {\em $2a$ shift rule} to indicate the range $2a$ of making the shift. 

In the following discussion, it will become clear that the fundamental set of Stokes matrices in the above sense is 
$\bigl\{ S_n \bigr\}_{n=0}^{2r-1}$, therefore essentially a single {\em symmetric Stokes matrices}%
\footnote{The name comes from the symmetric Stokes matrices in the $\mathbb Z_k$-symmetric critical points \cite{CIY2}, although there is no $\mathbb Z_k$-symmetry in the fractional-superstring critical points.} 
$S_0^{(\rm sym)}$. Here the symmetric Stokes matrices are defined as 
\begin{align}
S_{2rn}^{(\rm sym)} \equiv S_{2rn} S_{2rn+1} \cdots S_{2r(n+1)-1},\qquad \bigl(n=0,1,2,\cdots,k-1\bigr). 
\label{DefinitionOfSymStokesMatrices}
\end{align}
Therefore, we will start with {\em the symmetric Stokes sectors} $\bigl\{D_{2rn}\bigr\}_{n=0}^{k-1}$ 
in the following discussion of the multi-cut boundary conditions. 

\subsection{The multi-cut BC recursion equations}

\subsubsection{The boundary conditions and the shift rule}

The scaling orthonormal polynomial $\Psi_{\rm orth}(t;\zeta)$ 
of Eq.~\eq{ScalingOrthonormalPolynomials} is generally expressed 
by the canonical solution $\widetilde \Psi_{2rn}(t;\zeta)$ of the symmetric Stokes sectors $D_{2rn}$ as 
\begin{align}
\Psi_{\rm orth}(t;\zeta)= \widetilde \Psi_{2rn}(t;\zeta) \, X^{(2rn)}, \qquad 
X^{(2rn)} = 
\begin{pmatrix}
x_1^{(2rn)} \cr x_2^{(2rn)} \cr \vdots \cr x_k^{(2rn)}
\end{pmatrix}, \label{DefinitionOfPsiOrthWithVectorsX}
\end{align}
for each index $n=0,1,2,\cdots,k-1$. Therefore, the vectors $X^{(2rn)}$ are related 
by symmetric Stokes matrices of Eq.~\eq{DefinitionOfSymStokesMatrices} as
\begin{align}
X^{(2rn)} = S_{2rn}^{(\rm sym)}\, X^{(2r(n+1))},\qquad \bigl(n=0,1,2,\cdots,k-1\bigr). 
\label{MCBCRecursionVectorNotation}
\end{align}
Then the multi-cut boundary condition is imposed on the vectors $\bigl\{X^{(2rn)}\bigr\}_{n=0}^{k-1}$. 
By following the procedure discussed in \cite{CIY2}, 
we read the boundary conditions as follows: 
Here we first consider an example of $7$-cut $r=3$ $\omega^{1/2}$-rotated case. 
The profile in $D_0$ and the corresponding vector $X^{(0)}$ are given as 
\begin{align}
D_0  \supset \left[
\begin{array}{c|c|c|c|c|c|c}
1 & 7 & 2 & 6 & 3 & 5 & 4^\triangle \cr
\hline
7 & 1 & 6 & 2 & 5 & 3 & 4^\triangle \cr
\hline
7 & 6 & 1 & 5 & 2 & 4^\triangle & 3^\times \cr
\hline
6 & 7 & 5 & 1 & 4^\triangle & 2^\times & 3^\times \cr
\hline
6 & 5 & 7 & 4 & 1^\triangle & 3^\times & 2^\times \cr
\hline
5 & 6 & 4 & 7 & 3 & 1^\triangle & 2^\times \cr
\hline
5 & 4 & 6 & 3 & 7 & 2 & 1^\triangle \cr
\hline
4 & 5 & 3 & 6 & 2 & 7 & 1^\triangle 
\end{array}
\right],\qquad 
X^{(0)}= 
\begin{pmatrix}
x_1^{(0)}\neq 0 \cr 0 \cr 0 \cr x_4^{(0)} \neq 0 \cr x_5^{(0)} \cr x_6^{(0)} \cr x_7^{(0)}
\end{pmatrix}. 
\label{ProfileIn7CutR3OmegaWithMCBC}
\end{align}
Note that "$2^\times$" means $x_2^{(0)}=0$ and ``$4^\triangle$'' means $x_4^{(0)}\neq 0$. 
In the middle of the above profile, there is a horizontal line where ``$1^\triangle$'' and ``$4^\triangle$'' change their dominance, which is interpreted as a physical cut of the resolvent operator. 
In principle we should perform the same procedure 
for all the symmetric Stokes sectors $\bigl\{D_{2rn}\bigr\}_{n=0}^{k-1}$ 
to obtain their vectors $\bigl\{X^{(2rn)}\bigr\}_{n=0}^{k-1}$, however the boundary conditions 
for all the vectors $\bigl\{X^{(2rn)}\bigr\}_{n=0}^{k-1}$ are automatically 
obtained from the first vector $X^{(0)}$ in the first symmetric Stokes sector $D_0$ 
because {\em the multi-cut boundary conditions for the vectors 
$\bigl\{X^{(2rn)}\bigr\}_{n=0}^{k-1}$ are related to each other 
by the $2r$ shift rule of Eq.~\eq{ProfileShiftRule}:}
\begin{align}
x_j^{(2rn)}\neq 0\quad \Bigl( \text{or }\, x_j^{(2rn)}=0 \Bigr)\quad \Rightarrow\quad 
x_{j+rm_1}^{(2rn+2r)}\neq 0\quad \Bigl( \text{or }\, x_{j+rm_1}^{(2rn+2r)}=0\Bigr).
\label{VectorXShiftRule}
\end{align}
For example, one can directly check the other vectors $\bigl\{X^{(6n)}\bigr\}_{n=1}^6$ 
of this case ($k=7, r=3, m_1=1$) are given as 
\begin{align}
X^{(6)}= 
\begin{pmatrix}
x_1^{(6)} \cr x_2^{(6)} \cr x_3^{(6)} \cr 
x_4^{(6)}\neq 0 \cr 0 \cr 0 \cr x_7^{(6)} \neq 0 
\end{pmatrix},\qquad 
X^{(12)}= 
\begin{pmatrix}
0 \cr 0 \cr x_3^{(12)} \neq 0 \cr
x_4^{(12)} \cr x_5^{(12)} \cr x_6^{(12)} \cr 
x_7^{(12)}\neq 0 
\end{pmatrix},\qquad 
X^{(18)}= 
\begin{pmatrix}
x_1^{(18)} \cr x_2^{(18)} \cr 
x_3^{(18)}\neq 0 \cr 
0 \cr 0 \cr x_6^{(18)} \neq 0 \cr
x_7^{(18)} 
\end{pmatrix}\cdots,
\end{align}
and they are all consistent with Eq.~\eq{VectorXShiftRule}. 
This is because we read the multi-cut boundary conditions from the profiles which 
are again obtained by using the $2r$ shift rule. 
Therefore, only the multi-cut boundary condition equation, Eq.~\eq{MCBCRecursionVectorNotation}, 
of $n=0$ is essential and then the other equations are generated 
by using the $2r$ shift rule, Eq.~\eq{ProfileShiftRule}. 
This procedure reduces a great deal of the complexity. 
Note that this consideration does not mean there is the $\mathbb Z_k$ symmetry in the critical points. 
An analogy of this consideration is ``a symmetry of equation of motion%
\footnote{Maybe of non-critical M theory.},'' 
which does not necessarily mean a symmetry of the solutions. 

It is not difficult to read the boundary condition for general $k$ with $r=3$, 
which will be used in the practical analysis. 
We put $k=2r\widetilde m+\widetilde l$ $(1\leq \widetilde l\leq 2r)$ and there are $2r\,(=6)$ different cases 
($\widetilde l=1,2,\cdots,6$, respectively):
\begin{align}
\text{1. }& X^{(0)} \leftrightarrow (1^\triangle, \times,\times,4^\triangle,\times,\times, 7^\triangle,\cdots,\times,\times,
\bigl(3\widetilde m+1\bigr)^\triangle,--\cdots-); \nn\\
\text{2. }&X^{(0)} \leftrightarrow (1^\triangle,\times,\times,4^\triangle,\times,\times,7^\triangle,\cdots,\times,\times,
\bigl(3\widetilde m+1\bigr)^\triangle,\times,--\cdots-);
\nn\\
\text{3. }&X^{(0)} \leftrightarrow (1^\triangle, \times,\times,4^\triangle,\times,\times,7^\triangle,\cdots,\times,\times,
\bigl(3\widetilde m+1\bigr)^\triangle,\times,\times,--\cdots-);
\nn\\
\text{4. }&X^{(0)} \leftrightarrow (1^\triangle, \times,\times,4^\triangle,\times,\times,7^\triangle,\cdots,\times,\times,
\bigl(3\widetilde m+1\bigr)^\triangle,\times,\times,\bigl(3\widetilde m+4\bigr)^\triangle,--\cdots,-);
\nn\\
\text{5. }&X^{(0)} \leftrightarrow (1^\triangle, \times,\times,4^\triangle,\times,\times,7^\triangle,\cdots,\times,\times,
\bigl(3\widetilde m+1\bigr)^\triangle,\times,\times,\bigl(3\widetilde m+4\bigr)^\triangle,--\cdots,--);
\nn\\
\text{6. }&X^{(0)} \leftrightarrow (1^\triangle, \times,\times,4^\triangle,\times,\times,7^\triangle,\cdots,\times,\times,
\bigl(3\widetilde m+1\bigr)^\triangle,\times,\times,\bigl(3\widetilde m+4\bigr)^\triangle,--\cdots,---). 
\label{MCBCofXforGeneralKWithR3}
\end{align}
To save the space, we used a simpler notation: ``$1^\triangle$'' means $x_1^{(0)}\neq 0$, 
``$\times$'' on the $i$-th 
component means $x_i^{(0)}=0$, and ``$-$'' means 
that there is no constraint on these components. 
For example, the vector $X^{(0)}$ of Eq.~\eq{ProfileIn7CutR3OmegaWithMCBC} is expressed as 
$(1^\triangle, \times,\times,4^\triangle,---)$. 
Here we introduced a notation of the indices with tildes, $(\widetilde m,\widetilde l)$. 
It is because this $2r$ classification is reduced to be a simpler $r$ classification, 
$k=rm+l$ $(1\leq l\leq r)$, in the final unified form of the multi-cut BC recursion equations.%
\footnote{Therefore, $1.\leftrightarrow 4.$, $2.\leftrightarrow 5.$, $3.\leftrightarrow 6.$ in Eq.~\eq{MCBCofXforGeneralKWithR3} have the same form of the multi-cut BC recursions. }

Finally we mention the meaning of the general $2a$ shift rule for the boundary condition. 
We used the $2r$ shift rule because the multi-cut boundary condition 
Eq.~\eq{ProfileIn7CutR3OmegaWithMCBC} 
(generally mentioned around Eq.~\eq{DefinitionOfChiAngleInMCBC}) 
only respects the $2r$ shift rule. 
Therefore, one may wonder if we use the other $2a$ shift rules on these multi-cut boundary conditions. 
The answer to this question is following: {\em The $2a$ shift rules ($a=1,2,\cdots,r-1$) 
for the multi-cut boundary conditions generate the complementary boundary conditions. }
The complementary boundary condition on the Stokes phenomena 
was first discussed in the $\mathbb Z_k$ symmetric cases \cite{CIY2}. 
The physical meaning of the complementary boundary conditions 
is still unclear but these constraints provide strong equations which help us 
obtain explicit expressions of the Stokes multipliers even in the general $(k,r)$ isomonodromy systems.%
\footnote{One may expect that it is a constraint on operators in non-critical M theory which is analogous to 
``degenerate operators in conformal field theory''. Also note that, if the indices $(k,r)$ are small integers, then 
the complementary boundary conditions are exactly satisfied. See \cite{CIY2} for explicit analysis of 
some concrete examples. }
In fact, one can check that the $2a$ shift rule ($a=1,2$) 
for the the multi-cut boundary condition Eq.~\eq{ProfileIn7CutR3OmegaWithMCBC}  
results in the following two kinds of complementary boundary conditions: 
\begin{align}
\text{2 shift:}\quad
&D_2  \supset \left[
\begin{array}{c|c|c|c|c|c|c}
2 & 1 & 3 & 7 & 4 & 6 & 5^\triangle \cr
\hline
1 & 2 & 7 & 3 & 6 & 4 & 5^\triangle \cr
\hline
1 & 7 & 2 & 6 & 3 & 5^\triangle & 4^\times \cr
\hline
7 & 1 & 6 & 2 & 5^\triangle & 3^\times & 4^\times \cr
\hline
7 & 6 & 1 & 5 & 2^\triangle & 4^\times & 3^\times \cr
\hline
6 & 7 & 5 & 1 & 4 & 2^\triangle & 3^\times \cr
\hline
6 & 5 & 7 & 4 & 1 & 3 & 2^\triangle \cr
\hline
5 & 6 & 4 & 7 & 3 & 1 & 2^\triangle 
\end{array}
\right],\qquad 
X^{(2)}= 
\begin{pmatrix}
x_1^{(2)}\cr x_2^{(2)}\neq 0 \cr 0 \cr 0 \cr x_5^{(2)} \neq 0 \cr x_6^{(2)} \cr x_7^{(2)} 
\end{pmatrix}; \\
\text{4 shift:}\quad 
&D_4  \supset \left[
\begin{array}{c|c|c|c|c|c|c}
3 & 2 & 4 & 1 & 5 & 7 & 6^\triangle \cr
\hline
2 & 3 & 1 & 4 & 7 & 5 & 6^\triangle \cr
\hline
2 & 1 & 3 & 7 & 4 & 6^\triangle & 5^\times \cr
\hline
1 & 2 & 7 & 3 & 6^\triangle & 4^\times & 5^\times \cr
\hline
1 & 7 & 2 & 6 & 3^\triangle & 5^\times & 4^\times \cr
\hline
7 & 1 & 6 & 2 & 5 & 3^\triangle & 4^\times \cr
\hline
7 & 6 & 1 & 5 & 2 & 4 & 3^\triangle \cr
\hline
6 & 7 & 5 & 1 & 4 & 2 & 3^\triangle 
\end{array}
\right],\qquad 
X^{(4)}= 
\begin{pmatrix}
x_1^{(4)} \cr x_2^{(4)}\cr x_3^{(4)}\neq 0 \cr 0 \cr 0 \cr x_6^{(4)} \neq 0 \cr x_7^{(4)} 
\end{pmatrix}. 
\label{ProfileIn7CutR3OmegaWithCompBC}
\end{align}
Therefore, in general, each of these $2a$ complementary boundary conditions $(a=1,2,\cdots,r-1)$ requires 
that there is a solution of the Baker-Akhiezer function system Eq.~\eq{DESystem} 
which has the physical cuts along the $k$ directions, $\zeta\to\infty \times e^{i\chi_n}$, of 
\begin{align}
\chi_n=\chi_0 + \frac{2\pi n}{k}, \qquad \chi_0 = \frac{\pi}{k}(1-\epsilon)+ \frac{2\pi a}{rk},\qquad 
\bigl(n=0,1,2,\cdots,k-1\bigr). \label{DefinitionOfChiAngleInCompBC}
\end{align}
In the following, it is also convenient not to distinguish the multi-cut boundary condition and 
the $2a$ complementary boundary conditions 
(defined by Eq.~\eq{DefinitionOfChiAngleInCompBC}), 
therefore, they are equally called {\em the multi-cut $2a$ boundary condition}. 
In the $a=0$ case, we skip $2a$ in the name. 

\subsubsection{The theta parametrization}

\begin{figure}[t]
 \begin{center}
  \includegraphics[scale=1]{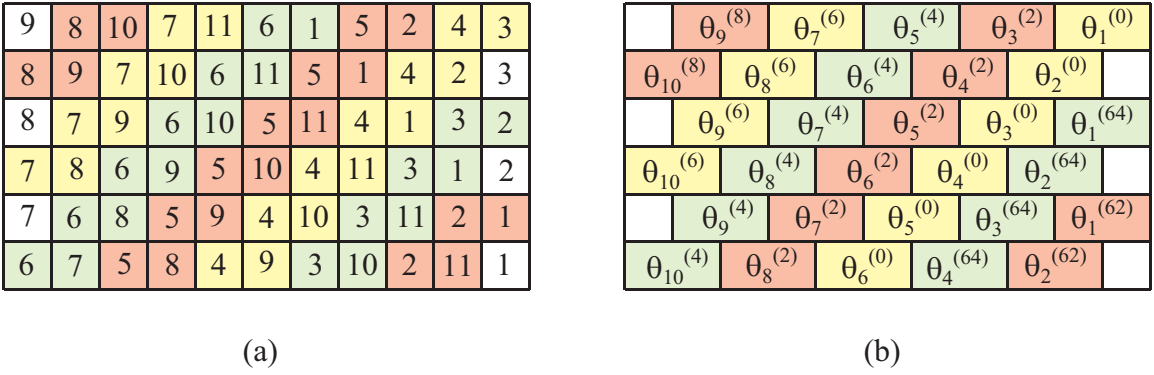}
 \end{center}
 \caption{\footnotesize 
The theta parametrization of fine Stokes multipliers in the $11$-cut ($r=3$) $\omega^{1/2}$-rotated models. 
Here only the fundamental domain $\bigcup_{l=0}^{2r-1} J_l$ of the profile is shown. 
a) The profile of dominant exponents with classification by different $3\, (=r)$ colors. Here we skipped 
the parenthesis notation $(i|j)$ but each pair of two same-colored boxes next to each other is 
the parenthesis pair $(i|j)$. b) The theta parametrization 
$\theta_n^{(2a)}$ is shown by using the correspondence $s_{l,i,j}\leftrightarrow (j|i)_l\in J_l$. 
The colors are drawn in accordance with $a$ modulo $r$. One can see the tilting pattern 
of the parametrization. 
} 
 \label{FigureThetaProfile}
\end{figure}

In the analysis of the $\mathbb Z_k$-symmetric critical points \cite{CIY2}, 
it was observed that the multi-cut boundary condition equation 
Eq.~\eq{MCBCRecursionVectorNotation} with the multi-cut boundary condition 
Eq.~\eq{DefinitionOfChiAngleInMCBC}
only includes a particular sequence of Stokes multipliers. 
Motivated by this observation given in \cite{CIY2}, 
here we also introduce the following parametrization of 
the fine Stokes multipliers: 
\begin{align}
\theta^{(2a+\epsilon r)}_{n} & \equiv  
s_{2r +2a -n+1+\epsilon r,\ds j_{2r +2a -n+1+\epsilon r, k-n+1}^{(\epsilon)},j_{2r + 2a -n+1+\epsilon r, k-n}^{(\epsilon)}},\nn\\
&\qquad\qquad \qquad   \bigl(n=1,2,\cdots,k-1;a=0,1,\cdots, rk-1 \bigr). 
\label{DefinitionOfThetaParameters}
\end{align}
It is much easier to catch the meaning of the formula by drawing it on the profile. 
Figure \ref{FigureThetaProfile} is an example of the profile for the $11$-cut ($r=3$) $\omega^{1/2}$-rotated models. 

In fact, the multi-cut BC recursion equations Eq.~\eq{MCBCRecursionVectorNotation} 
with the multi-cut boundary condition only include a subset of the Stokes multipliers, 
$\bigl\{\theta_n^{(2rb)}\bigr\}_{1 \leq n\leq k-1}^{0\leq b\leq k-1}$, as we expected. 
Importantly, the $2a$ shift rule acts 
on the parameter $\theta_n^{(2rb)}$ as
\begin{align}
\text{$2a$ shift:}\qquad 
\theta_n^{(2rb)}\quad 
\to \quad 
\theta_n^{(2rb+2a)}, 
\end{align}
therefore the multi-cut BC recursion equations Eq.~\eq{MCBCRecursionVectorNotation} with 
the multi-cut $2a$ boundary condition ($a=1,2,\cdots,r-1$) include a complementary set of 
the Stokes multipliers, 
$\bigl\{\theta_n^{(2rb+2a)}\bigr\}_{1 \leq n\leq k-1}^{0\leq b\leq k-1}$. 

Before we discuss the details of the multi-cut BC recursion equations, 
basic properties of the theta parameters are summarized: 

\begin{itemize}
\item The $\mathbb Z_2$-symmetry condition ($k\in 2\mathbb Z$) 
Eq.~\eq{Z2SymmetryCondition} 
is expressed as  
\begin{align}
\theta_n^{(2a)} = \theta_n^{(2a+rk)}\qquad \bigl(n=1,2,\cdots,k-1; \, a=0,1,2,\cdots, rk-1\bigr). 
\end{align}
\item The hermiticity condition Eq.~\eq{HermiticityCondStokes} is expressed as 
\begin{align}
\Bigl[\theta_n^{(a)}\Bigr]^*= - \theta_{k-n}^{(-4r-a)} 
\qquad \bigl(n=1,2,\cdots,k-1; \, a=0,1,2,\cdots, rk-1\bigr). 
\label{HermiticityConditionSigmaParameters}
\end{align}
\item For later convenience, we extend the parameter as 
\begin{align}
\theta_0^{(2a)} = -1,\qquad \theta_k^{(2a)} = +1,
\qquad \bigl(a=0,1,2,\cdots, rk-1\bigr),\label{Theta0k}
\end{align}
which is consistent with the hermiticity constraint. 
\end{itemize}

\subsubsection{The multi-cut BC recursion equations}

In this section, we discuss the multi-cut boundary condition equation Eq.~\eq{MCBCRecursionVectorNotation} 
with the multi-cut $2a$ boundary condition Eq.~\eq{DefinitionOfChiAngleInCompBC}. 
The derivation is given in Appendix \ref{SketchOfProofTheoremMCBCRE}, 
and therefore we here only show the result: 
\begin{Theorem}
[The multi-cut BC recursion equations]
In the fractional-superstring critical points of an arbitrary $(k,r)$ 
with ${\rm g.c.d.}\,(k,\gamma)=1$ ($\gamma=r-2$), 
the multi-cut boundary condition equations Eq.~\eq{MCBCRecursionVectorNotation}  
with the multi-cut $2a$ boundary condition Eq.~\eq{DefinitionOfChiAngleInCompBC} 
of $a=0,1,\cdots,r-1$ 
result in the following recursion equations: 
\begin{align}
&\mathcal F_b^{(2s)}\Bigl[\bigl\{ y_n^{(2s)}\bigr\}_{n\in \mathbb Z}\Bigr]
\equiv \sum_{
0 \leq r(n-1) + b \leq k,\, 
n \in \mathbb Z
}
\theta_{r(n-1)+b}^{\bigl(2r(n-1)+2s\bigr)} \, y_n^{(2s)} = 0 \nn\\
&\qquad \text{of }\qquad s=rl+a + \epsilon \frac{r}{2};\qquad  \bigl(0 \leq l\leq k-1,\quad 1\leq b \leq r\bigr), 
\label{GeneralFormulaForTheMultiCutBCRecursionEquation}
\end{align}
where the sequence of numbers $\bigl\{ y_n^{(2s)}\bigr\}_{n\in \mathbb Z}$ are all non-zero 
and satisfy
\begin{align}
y_n^{(2s)} = y_{n-b}^{(2s+2rb)},
\qquad y_{n}^{(2s)}= y_{n+k}^{(2s)},
\qquad \bigl(n,a,b\in \mathbb Z\bigr). 
\end{align}
Therefore, there are $r$ distinct sequences 
$\bigl\{ y_{n}^{(2a)}\bigr\}_{n=1}^k$ corresponding to each boundary condition of $2a$. 
\label{TheoremMultiCutBCRecursionEquations}
\end{Theorem}
For each boundary condition (of $2a$) there are $r$ recursive equations 
$\mathcal F_b^{(2s)}=0$ of $b=1,2,\cdots,r$. It is suggestive to see 
the positions of the Stokes multipliers (included in each equation $\mathcal F_b^{(2s)}=0$) 
in the profile. Here we do not show explicitly, but one can see that the trajectory is along 
the line vertical to the trajectories of colors in Figure \ref{FigureThetaProfile}. 

Although the formula Eq.~\eq{GeneralFormulaForTheMultiCutBCRecursionEquation} 
is simply written in a unified way, 
the concrete form of each equation system depends on $k$ of modulo $r$: $k=rm+l$. 
Here we show each of them. First of all, we introduce a new parameter
\begin{align}
\sigma_{m-n+1,b}^{(2a)} \equiv \theta_{r(n-1)+b}^{(2r(n-1)+2a)},\qquad 
m\equiv \Bigl\lfloor \frac{k-1}{r} \Bigr\rfloor,
\end{align}
and then the multi-cut BC recursion equations are 
\begin{align}
\mathcal F_b^{(2s)}\Bigl[\bigl\{ y_n^{(2s)}\bigr\}_{n\in \mathbb Z}\Bigr]
\equiv \sum_{n=\lceil \frac{r-b}{r}\rceil}^{\lfloor \frac{k+r-b}{r} \rfloor}
\sigma_{m+1-n,b}^{(2s)} \, y_n^{(2s)} = 0.
\label{GeneralFormulaForTheMultiCutBCRecursionEquationInTermsOfSigma}
\end{align}
The concrete form of each equation is shown in Table \ref{TableClassificationOfBCRE}. 
Here we also show the hermiticity condition 
Eq.~\eq{HermiticityConditionSigmaParameters} in terms of the sigma parameter: 
\begin{align}
\sigma_{m-n+1,b}^{(2s)} = 
\left\{
\begin{array}{ll}
-\Bigl[\sigma_{n-1,(l-b)}^{(-2r(m+2)-2s)}\Bigr]^* & : 1\leq b < l, 
\quad \bigl(n=1,2,\cdots,m+1\bigr) \cr
-\Bigl[\sigma_{n,(r+l-b)}^{(-2r(m+1)-2s)}\Bigr]^* & : l\leq b \leq r, 
\quad \bigl(n=1,2,\cdots,m\bigr)
\end{array}
\right.. 
\label{HermiticityConditionInSigma}
\end{align}
Therefore, the hermiticity condition relates equations of the five cases (I, II, III, IV,V) as: 
\begin{align}
{\rm I} \leftrightarrow {\rm I},\qquad 
{\rm II} \leftrightarrow {\rm IV},\qquad 
{\rm III} \leftrightarrow {\rm III},\qquad
{\rm V} \leftrightarrow {\rm V}. 
\end{align}
These are consistent 
with Eq.~\eq{GeneralFormulaForTheMultiCutBCRecursionEquationInTermsOfSigma}
(also see Table \ref{TableClassificationOfBCRE}).

\begin{table}[t]
\begin{center}
\begin{tabular}[c]{c|c|c|c}
\hline
cases & 
$1\leq b\leq r$ & $\ds \mathcal F_b^{(2s)}\bigl[\bigl\{ y_n^{(2s)}\bigr\}_{n\in\mathbb Z}\bigr]=0$ & length \\
\hline \hline
I. & $1\leq b <l$ & $\ds \sum_{n=1}^{m+1}\sigma_{m-n+1,b}^{(2s)}\, y_n^{(2s)}=0$ & $m+1$ \\
\hline
II. & $b=l \,(\neq r)$ & $\ds y_{m+1}^{(2s)} + \sum_{n=1}^{m}\sigma_{m-n+1,b}^{(2s)}\, y_n^{(2s)}=0$ & $m+1$ \\
\hline
III. & $l<b<r$ & $\ds \sum_{n=1}^{m}\sigma_{m-n+1,b}^{(2s)}\, y_n^{(2s)}=0$ & $m$ \\
\hline
IV. & $b=r\, (\neq l)$ & $\ds -y_0^{(2s)} + \sum_{n=1}^{m}\sigma_{m-n+1,b}^{(2s)}\, y_n^{(2s)}=0$ & $m+1$ \\
\hline \hline
V. & $b=r\, (= l)$ & $\ds -y_0^{(2s)} +y_{m+1}^{(2s)}+ \sum_{n=1}^{m}\sigma_{m-n+1,b}^{(2s)}\, y_n^{(2s)}=0$ & $m+2$ \\
\hline
\end{tabular}
\end{center}
\caption{\footnotesize The concrete forms of 
Eq.~\eq{GeneralFormulaForTheMultiCutBCRecursionEquation} 
(and then Eq.~\eq{GeneralFormulaForTheMultiCutBCRecursionEquationInTermsOfSigma}) 
is shown with $k=rm+l,\, (1\leq l\leq r)$. 
The case of V is the degenerate case of II and IV (i.e.~$r=l$). 
Possible cases are (II, III, IV) for $l=1$, (I, II, III, IV) for $1<l<r$, and (I, V) for $l=r$. 
The ``length'' means the number of terms in the equation 
$\mathcal F_b^{(2s)}\bigl[\bigl\{ y_n^{(2s)}\bigr\}_{n\in\mathbb Z}\bigr]=0$. }
\label{TableClassificationOfBCRE}
\end{table}

\subsection{Relation to discrete Hirota dynamics \label{SubsectionHirota}}

Here we point out an intriguing connection to the structure of quantum integrable models. 
A non-trivial connection between 
Stokes phenomena in ordinary differential equations 
and T-systems of quantum integrable systems is found in a special kind of Schr\"odinger 
equations \cite{ODEIMRef,JSuzuki}, which is called {\em the ODE/IM correspondence}.%
\footnote{The authors would like to thank Toshio Nakatsu and Ryo Suzuki 
for drawing our attention to these interesting papers and the correspondence. } 
In this section, we provide a new generalization of this correspondence to 
general ODE systems appearing in the isomonodromy theory. 

In general, the Stokes multipliers 
are distributed in the Stokes matrices $\bigl\{S_n\bigr\}_{n=0}^{2rk-1}$ in a very complicated way. 
However, if one looks at the general multi-cut BC equations 
Eq.~\eq{GeneralFormulaForTheMultiCutBCRecursionEquation}, 
one notices that the Stokes multipliers are re-organized in the following suggestive form: 
For example, the equation of IV in Table \ref{TableClassificationOfBCRE} is expressed as 
\begin{align}
Y^{(2s-2r)} = \Phi^{(2s)}\, Y^{(2s)},\qquad \bigl(s=0,1,2,\cdots,rk-1\bigr),  
\label{JSrecursion}
\end{align}
with 
\begin{align}
Y^{(2s)} \equiv 
\begin{pmatrix}
y_1^{(2s)} \cr 
y_2^{(2s)} \cr
\vdots \cr
y_m^{(2s)}
\end{pmatrix},
\qquad 
\Phi^{(2s)} = 
\begin{pmatrix}
\tau_1^{(2s)} & \tau_{2}^{(2s)} & \cdots & \tau_{m-1}^{(2s)} & \tau_m^{(2s)} \cr
1 & 0 &  \cdots & 0 &  0 \cr
0   & 1 &  & 0 & 0 \cr
\vdots   &    & \ddots & \vdots & \vdots \cr
 0  &  0  & \cdots            & 1          & 0 
\end{pmatrix}.
\label{JSrecursionContents}
\end{align}
Here $\bigl\{\tau_n^{(2s)}\bigr\}_{n\in\mathbb Z}$ are related to 
the fine Stokes multipliers $\bigl\{\sigma_{n,b}^{(2s)}\bigr\}_{n\mathbb Z}$ as
\begin{align}
\tau_n^{(2s)} = \sigma_{m-n+1,r}^{(2s)},\qquad \bigl(n=1,2,\cdots,m\bigr). 
\label{TauSigmaIV}
\end{align}
This is the same form of the recursive equations found 
in the study of Stokes phenomena for 
the Schr\"odinger-equation systems of the ODE/IM correspondence (See \cite{JSuzuki} for example). 
Therefore, according to the discussion in \cite{JSuzuki}, each sequence 
$\bigl\{\tau_n^{(2rl + 2b)}\bigr\}_{1\leq n\leq m}^{1\leq l\leq k}$ ($b=1,2,\cdots,r)$ satisfies 
discrete Hirota equations \cite{HirotaEqRef} (or a T-system) of quantum integrable systems.%
\footnote{For references of Hirota equations, see \cite{TsysRef} for example. }
An important difference from the original correspondence is that {\em in the $k\times k$ isomonodromy ODE system 
of the Poincar\'e index $r$, there are $r$ different recursive systems of Stokes multipliers which result 
in $r$ different T-systems}. That is, the system is described by 
{\em multiple T-systems of quantum integrable systems}. 

We note that, in our calculation, 
there is no explicit appearance of {\em spectral parameters} of quantum integrable systems. 
However the spectral parameters appear by assuming {\em analyticity of the index $2s$}. 
That is, we rewrite the index $s$ as $s=ru + a$ ($0\leq a\leq r-1$) and 
express them as follows: 
\begin{align}
\tau^{(2s)}_n = \tau_n (u;a),\qquad y_n^{(2s)} = y_0(u+n;a),\qquad Y^{(2s)} = Y(u;a). 
\end{align}
Then the claim is that analytic continuation of the index $u$ 
realizes the spectral parameters of the T-systems. This is understood as continuous deformation 
of the solutions for Stokes multipliers. Because of the isomonodromy property, 
therefore, this spectral parameter $u$ commutes with 
the original spectral parameter $\zeta$ of the resolvent operator: 
\begin{align}
\Bigl[\frac{\del}{\del u},\frac{\del}{\del \zeta}\Bigr] \Psi(t;\zeta;u) = 0. 
\end{align}
In this sense, this is a new spectral parameter which describes 
the strong-coupling dynamics of non-critical string theory, possibly, non-critical M theory. 

\subsubsection{Hirota dynamics for non-critical string/M theory}

\begin{figure}[t]
 \begin{center}
  \includegraphics[scale=0.8]{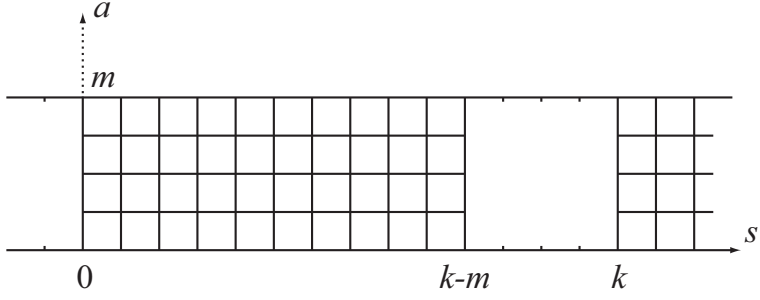}
 \end{center}
 \caption{\footnotesize 
The boundary condition for the T-system with $U_q(A_{m-1}^{(1)})$ of $q^k=1$. 
The lattice points explicitly shown in the figure are the support of the T-function, $T_{a,s}(u)$. 
In particular, $T_{0,s}(u)=T_{a,0}(u)=1\, (s\in \mathbb Z, 0\leq a \leq m$), 
and also there is a periodicity $T_{a,s}(u)=T_{a,s+k}(u)$. 
The example shown here is $k=14$ and $m=4$ (i.e.~$r=3$). 
} 
 \label{FigureTsystemBC}
\end{figure}

Here we skip ``$a$'' which does not change in each integrable model. 
The direct consequence of \cite{JSuzuki} is following: 
T-system appears as the algebraic relations encoded in the recursion equation 
Eq.~\eq{JSrecursion} and Eq.~\eq{JSrecursionContents}, 
i.e.~in the algebraic relations among components 
$\bigl\{\phi_{n,i,j}\bigr\}_{1\leq i,j\leq m}^{n\in\mathbb Z}$ of the multiplication matrices 
$\Phi_n(u) = \bigl(\phi_{n,i,j}(u)\bigr)_{1\leq i,j\leq m}$: 
\begin{align}
\Phi_n(u)\equiv 
\Phi \bigl(u\bigr)\, \Phi \bigl(u+1\bigr)\cdots \Phi \bigl(u+n-2\bigr)\, \Phi \bigl(u+n-1\bigr). 
\label{DefinitionPhiNJS}
\end{align}
The discrete Hirota equation (i.e.~T-system) which is relevant to the algebraic structure 
is that of the integrable models with the quantum symmetry $U_q(A_{m-1}^{(1)})$ of $q^k=1$ \cite{JSuzuki}: 
\begin{align}
T_{a,s}(u+1)\, T_{a,s}(u-1) = T_{a,s+1}(u)\, T_{a,s-1}(u) + T_{a+1,s}(u) \, T_{a-1,s}(u),
\end{align}
with the following boundary condition: 
\begin{align}
&T_{-n,s}(u)=T_{m+n,s}(u)  =0 \qquad \bigl(n \in \mathbb N,\, s\in \mathbb Z\bigr),\label{TsystemBC1}\\
&T_{a,s+k-m} (u)=0\qquad \bigl(0< a <m, \, 0 < s < m\bigr), \label{TsystemBC2}\\
&T_{a,0}(u)=T_{0,s}(u) =1\qquad (0 \leq a \leq m,\, s\in \mathbb Z), \label{TsystemBC3}
& T_{a,s}(u)=T_{a,s+k}(u), 
\end{align}
which is also shown in Figure \ref{FigureTsystemBC}. 
Note that T-functions also satisfy periodicity in the spectral parameter $u$: 
\begin{align}
T_{a,b}(u+2k) = T_{a,b}(u). 
\end{align}
The correspondence of T-functions with the components of the $\Phi_n(u)$ is given in the following way \cite{JSuzuki}:%
\footnote{The expression for more general components is also shown in \cite{JSuzuki} (i.e.~Theorem 3). }
\begin{align}
\tau_n(u) =(-1)^{n+1} T_{n,1}(2u+n), \qquad 
\phi_{n,1,1}(u) = T_{1,n}(2u+n). 
\label{TfunctionTauPhiIdentification}
\end{align}
Notes on the boundary conditions are in order: 
\begin{itemize}
\item [1.] Generally it is known that 
general T-functions of the above T-system can be expressed in terms of the 
vertical/horizontal inputs by the Bazhanov-Reshetikhin formula \cite{BRformula}
\begin{align}
T_{a,s}(u)& = \det_{1\leq i, j \leq s} T_{a+i-j,1}(u+s+1-i-j), \label{BRformula1}\\
&= \det_{1\leq i, j \leq a} T_{1,s+i-j}(u+a+1-i-j), \label{BRformula2}
\end{align}
with the boundary condition Eq.~\eq{TsystemBC3}. 
\item [2.] One can directly solve Eq.~\eq{DefinitionPhiNJS} and check that 
the coefficients $\bigl\{\phi_{n,1,1}(u)\bigr\}_{n=1}^\infty$ are expressed by $\bigl\{\tau_n(u)\bigr\}_{n=1}^m$ 
and the expression is equivalent to the Bazhanov-Reshetikhin formula of Eq.~\eq{BRformula1} 
(and the identification Eq.~\eq{TfunctionTauPhiIdentification})
with the following extension of an index: 
\begin{align}
\tau_{0}(u) = -1,\qquad \tau_{-n}(u) = \tau_{m+n}(u)=0\qquad \bigl(n\in\mathbb N\bigr). 
\end{align}
In particular, Eq.~\eq{BRformula1} results in Eq.~\eq{TsystemBC1} and Eq.~\eq{TsystemBC3}. 
\item [3.] The cyclic condition $\Phi_k(u)=I_m$ 
(which results from the monondromy free condition Eq.~\eq{OriginalMonodromyFreeCondition} in our case) 
gives the following constraints on the components $\bigl\{\phi_{n,1,1}(u)\bigr\}_{n\in \mathbb Z}$: 
\begin{align}
\phi_{n,1,1} = \phi_{n+k,1,1}\qquad \bigl(n\in\mathbb Z\bigr),\qquad 
\phi_{k-m+n,1,1} = \delta_{n,m}\qquad \bigl(1 \leq n\leq m\bigr). 
\label{MFconditionInPHI}
\end{align}
Therefore, by using Eq.~\eq{BRformula2}, one can show Eq.~\eq{TsystemBC2} and Eq.~\eq{TsystemBC3}. 
\end{itemize}

\subsection{Explicit solutions of Stokes multipliers \label{SubsectionExplicitSolutions}}

Here we show several explicit solutions for the Stokes multipliers 
which generalize the discrete solutions found in the $\mathbb Z_k$-symmetric critical points \cite{CIY2}. 
Solutions shown here are therefore special solutions and further analysis for the complete solutions 
(with using the above T-systems and their Bathe ansatz equations) 
should be studied in another publication. 
In particular, we impose the following ``$\mathbb Z_k$-symmetry condition'': 
\begin{align}
&S_{n+2r} = \Gamma^{-1} S_n \Gamma \qquad \bigl(n=0,1,\cdots,2rk-1\bigr)\nn\\
\Leftrightarrow\qquad & \theta_n^{(2s)} = \theta_n^{(2s+2r)}\qquad \bigl(1\leq n\leq k-1;\, s=0,2,\cdots,rk-1\bigr). 
\end{align}
Note that this constraint on the Stokes multipliers does not necessarily mean 
existence of the $\mathbb Z_k$ symmetry of the system, 
although the monodromy free condition 
Eq.~\eq{OriginalMonodromyFreeCondition} becomes 
\begin{align}
\bigl[S_0^{(\rm sym)}\Gamma^{-1}\bigr]^k=I_k. 
\label{MonodromyFreeConditionZKSymmetric}
\end{align}
From the viewpoint of the discrete Hirota equation, 
the $\mathbb Z_k$-symmetric constraint is equivalent to dropping the spectral-parameter dependence 
and the Hirota equation then becomes the system for characters of a corresponding Lie group. 
Consequently, the Stokes multipliers $\bigl\{\tau_n(u)\bigr\}_{n=1}^m$ 
of Eq.~\eq{JSrecursionContents} are given 
by the characters of the anti-symmetry representations, $\chi_{n,1}(g)$: 
\begin{align}
\tau_n(u) = (-1)^{n+1} T_{n,1}(u+2n) = (-1)^{n+1} \chi_{n,1}(g), 
\end{align}
with a proper group element $g \in {\rm GL}(m)$. 
The constraints on the group element $g$ is from the algebraic relation of the T-system boundary conditions, 
however, we know the solutions since the system is the same as 
this $\mathbb Z_k$-symmetric situation \cite{CIY2}. That is, the group elements are given by 
distinct roots of unity, $\bigl\{\omega^{n_j}\bigr\}_{j=1}^m$ and the characters are 
given by symmetric polynomials: 
\begin{align}
\chi_{n,1}(g) ={\rm Sym}\Bigl[n,\bigl\{\omega^{n_i}\bigr\}_{i=1}^m\Bigr],\qquad 
g \simeq \diag\bigl(\omega^{n_1},\omega^{n_2},\cdots,\omega^{n_m}\bigr). 
\end{align}
This is the connection to the discrete solutions found in \cite{CIY2}. 

Clearly this is not the end of story because there are now $r$ different T-systems,
\begin{align}
\tau_n(u) \quad \to \quad  \tau_n(u;a) \qquad \bigl(a=0,1,2,\cdots,r-1\bigr), 
\end{align}
and they are 
related by the original monodromy free condition Eq.~\eq{OriginalMonodromyFreeCondition}. 
This condition can be solved by showing explicit $k$ distinct eigenvectors of the symmetric Stokes matrix 
$S_{0}^{(\rm sym)}$ \cite{CIY2}. 
This condition results in the constraints on the charges (or group elements) $g(a)$ 
which are classified by configurations of the avalanches \cite{CIY2}. 

\subsubsection{The case of $k=rm+1$}
In this case, we have three kinds of equations (II, III and IV in Table \ref{TableClassificationOfBCRE}). 
Therefore, we have the following recursion-equation systems 
\begin{align}
{\rm II}:\qquad & Y(u+1;a) = \widetilde \Phi (u;a)\, Y (u;a), \qquad \bigl(a=0,1,\cdots, r-1\bigr), \nn\\
{\rm III}:\qquad & 0 = v_b (u;a)\, Y (u;a), \qquad \bigl(1 < b < r;\, a=0,1,\cdots, r-1\bigr), \nn\\
{\rm IV}:\qquad & Y(u-1;a) = \Phi (u;a)\, Y (u;a), \qquad \bigl(a=0,1,\cdots, r-1\bigr). 
\label{JSrecursionSp}
\end{align}
The matrix $\Phi(u;a)$ is defined by Eq.~\eq{JSrecursionContents} and Eq.~\eq{TauSigmaIV}, and 
the others are defined as 
\begin{align}
\widetilde \Phi (u;a) &= 
\begin{pmatrix}
0& 1 &  0 & \cdots & 0 &  0 \cr
0   & 0 & 1 & \cdots &  0& 0 \cr
\vdots   &  \vdots  & \vdots & \ddots &\vdots & \vdots \cr
 0  &  0  & 0 &            & 1          & 0  \cr
 0  &  0  &0& \cdots            & 0          & 1  \cr
-\sigma_{m,1}^{(2s)} & -\sigma_{m-1,1}^{(2s)}& -\sigma_{m-2,1}^{(2s)}& \cdots & - \sigma_{2,1}^{(2s)}&  -\sigma_{1,1}^{(2s)}
\end{pmatrix},\nn\\
v_b(u;a)&=
\bigl(
\sigma_{m,b}^{(2s)}, \sigma_{m-1,b}^{(2s)}, \sigma_{m-2,b}^{(2s)}, \cdots,   \sigma_{2,b}^{(2s)}, 
\sigma_{1,b}^{(2s)}
\bigr)\qquad \bigl(1<b<r\bigr). 
\end{align}
We here require%
\footnote{Note that there are other solutions which does not satisfy this condition. 
For example, we can also consider recursion equation of type III and define T-system of $U_q(A_{m-2}^{(1)})$. 
Here however, we only focus on special solutions which are relatively simpler. }
\begin{align}
\widetilde \Phi (u;a) = \Phi^{-1} (u+1;a),\qquad v_b (u;a)=0,
\end{align}
then the recursion equations is given by a single equation Eq.~\eq{JSrecursion} discussed 
in Section \ref{SubsectionHirota}, which means that there are $r$ Hirota equations. 
In terms of the components, these Stokes multipliers are given as%
\footnote{Note that the inverse of the matrix $\Phi^{(2s)}$ is given as 
\begin{align}
\bigr[\Phi^{(2s)}\bigl]^{-1} = 
\begin{pmatrix}
0& 1 &  0 & \cdots & 0 &  0 \cr
0   & 0 & 1 & \cdots &  0& 0 \cr
\vdots   &  \vdots  & \vdots & \ddots &\vdots & \vdots \cr
 0  &  0  & 0 &            & 1          & 0  \cr
 0  &  0  &0& \cdots            & 0          & 1  \cr
 \frac{1}{\tau_m^{(2s)}} & \frac{-\tau_{1}^{(2s)}}{\tau_m^{(2s)}} 
& \frac{-\tau_{2}^{(2s)}}{\tau_m^{(2s)}}& \cdots &  \frac{-\tau_{m-2}^{(2s)}}{\tau_m^{(2s)}} &  \frac{-\tau_{m-1}^{(2s)}}{\tau_m^{(2s)}}
\end{pmatrix}.
\end{align}}
\begin{align}
{\rm II}:\qquad & 
\sigma_{m,1}^{(2s)} = \frac{-1}{\tau_m^{(2s)}},\qquad 
\sigma_{m-n,1}^{(2s)} = \frac{\tau_n^{(2s)}}{\tau_m^{(2s)}} \qquad \bigl(n=1,2,\cdots,m-1\bigr),\nn\\
{\rm III}:\qquad & 
\sigma_{n,b}^{(2s)} = 0 \qquad \bigl(1<b<r;\, n=1,2,\cdots,m\bigr), \nn\\
{\rm IV}:\qquad & 
\sigma_{m-n+1,r}^{(2s)} = \tau_n^{(2s)},\qquad \bigl(n=1,2,\cdots,m\bigr). 
\end{align}
As special solutions to the Hirota equation for 
$\bigl\{\tau_n(u;a)\bigr\}_{1\leq n \leq m}^{0\leq a\leq r-1}$ 
(discussed in the beginning of this section), 
we assign the following discrete solutions 
\begin{align}
\tau_n(u;a) = (-1)^{n+1}{\rm Sym}\Bigl[n,\bigl\{\omega^{n_i^{(2a)}}\bigr\}_{i=1}^m\Bigr]\qquad 
\bigl(a=0,1,\cdots,r-1\bigr). 
\label{DiscreteSolutionTauFormula}
\end{align}
Once we derive this formula, the procedure to obtain solutions is the same as the $\mathbb Z_k$-symmetric cases 
\cite{CIY2}. Here we explicitly checked the constraints by Mathematica, 
and therefore, we make the following conjecture for the conditions on the exponents of $\omega$ 
in Eq.~\eq{DiscreteSolutionTauFormula}: 
$\bigl\{n_j^{(2a)}\bigr\}_{1\leq j\leq m}^{0\leq a\leq r-1}$
\begin{align}
1.\qquad &n_i^{(2a)} \not\equiv n_j^{(2a)} \quad \mod k \qquad \bigl(i\neq j;\, 0\leq a\leq r-1\bigr) 
\label{ConstraintOnIndexN1} \\
2.\qquad & n_i^{(2a)} \not \equiv n_0\equiv -\sum_{\begin{subarray}{c} 1\leq j\leq m \cr 0\leq b\leq r-1\end{subarray}} n_j^{(2b)}+ \Bigl(\frac{k}{2}\Bigr)\frac{1+(-1)^k}{2}\quad \mod k\nn\\
&\qquad \qquad \qquad\qquad \qquad\qquad \qquad \bigl(1\leq i\leq m;\, 0\leq a\leq r-1\bigr).  
\label{ConstraintOnIndexN2} \\
3.\qquad & \bigl\{n_j^{(2a)}\bigr\}_{1\leq j\leq m} = \bigl\{n_j^{(2(r-a))}\bigr\}_{1\leq j\leq m} \qquad 
\bigl(0\leq a \leq r-1\bigr). 
\label{ConstraintOnIndexN3}
\end{align}
Some notes are in order: 
\begin{itemize}
\item In order to satisfy the monodromy free condition 
Eq.~\eq{MonodromyFreeConditionZKSymmetric}, one should show that 
the symmetric Stokes matrix $S_0^{(\rm sym)}$ is diagonalizable. 
With the explicit expression of the discrete solution Eq.~\eq{DiscreteSolutionTauFormula}, 
one can explicitly construct $rm \, (=k-1)$ eigenvectors $\mathcal X(a;n_j^{(2a)})$ \cite{CIY2}, 
\begin{align}
S_0^{(\rm sym)}\Gamma^{-1} \, \mathcal X(a;n_j^{(2a)}) 
= \omega^{n_j^{(2a)}}\,\mathcal X(a;n_j^{(2a)})
\qquad \bigl(1\leq j\leq m;\, 0\leq a\leq r-1\bigr). 
\end{align}
A brief summary of the procedure is following: 
Each T-system of $\bigl\{T_n(u;a)\bigr\}_{n=1}^m$ (for a fixed $a=0,1,\cdots,r-1$) 
admits $m$ different vectors $Y(u;a)$:
\begin{align}
Y(u;a) = \mathcal Y(u;a;n_j^{(2a)})\equiv 
\begin{pmatrix}
\omega^{-(u+1)n_j^{(2a)}} \cr \omega^{-(u+2)n_j^{(2a)}} \cr 
\vdots \cr
\omega^{-(u+m)n_j^{(2a)}}
\end{pmatrix}
\qquad \bigl(1\leq j\leq m\bigr),
\end{align}
which satisfy 
\begin{align}
\Phi(u;a) \, \mathcal Y(u;a;n_j^{(2a)})  = \omega^{n_j^{(2a)}}\, \mathcal Y(u;a;n_j^{(2a)}). 
\end{align}
By using the recursion equation Eq.~\eq{MCBCRecursionVectorNotation}, 
one can find the explicit expression of each eigenvector $\mathcal X(a;n_j^{(2a)})$ 
in terms of $\mathcal Y(u;a;n_j^{(2a)})$: 
\begin{align}
\mathcal X(a;n_j^{(2a)}) = M(u;a)\, \mathcal Y(u;a;n_j^{(2a)}),
\end{align}
with a $k\times m$ matrix $M(u;a)$ which satisfies 
\begin{align}
S_{0}^{(\rm sym)}\Gamma^{-1} M(u;a) = M(u-1;a) \Phi(u;a),
\end{align}
then one obtains the explicit expression for $rm$ vectors 
$\bigl\{\mathcal X(a;n_j^{(2a)})\bigr\}_{1\leq j\leq m}^{0\leq a\leq r-1}$. 
Since each multi-cut $2a$ boundary condition is distinct to each other, 
the vectors are also distinct if and only if the condition Eq.~\eq{ConstraintOnIndexN1} is satisfied. 
\item Although the $rm \, (=k-1)$ distinct eigenvectors are explicitly constructed in the above procedure, 
there remains one missing eigenvector. 
Noting that $\det \bigl[S_0^{(\rm sym)}\Gamma^{-1}\bigr] = (-1)^{k-1}$, 
one can conclude that the eigenvalue associated with this vector is given by $\omega^{n_0}$ of 
\begin{align}
n_0\equiv -\sum_{\begin{subarray}{c} 1 \leq j\leq m \cr 0\leq b\leq r-1\end{subarray}} 
n_j^{(2b)} + \Bigl(\frac{k}{2}\Bigr)\frac{1+(-1)^k}{2}\quad \mod k, 
\end{align}
and the eigenvector is given as
\begin{align}
\mathcal X(n_0) = \sum_{a=0}^{r-1} c_a \, M(u;a) \mathcal Y(u;a;n_0). 
\end{align}
It is clear and the constraint Eq.~\eq{ConstraintOnIndexN2} is sufficient. 
The necessity is non-trivial but we explicitly check it by Mathematica for several cases. 
\item 
These Stokes multipliers $\bigl\{\tau_n(u;a)\bigr\}_{1\leq n \leq k-1}^{0\leq a\leq r-1}$ 
should be subjected to the hermiticity condition Eq.~\eq{HermiticityConditionInSigma}. 
This constraint results in the constraint Eq.~\eq{ConstraintOnIndexN3}. 
\end{itemize}

\subsubsection{The case of $k=rm+l$, $(1<l<r)$}
In this case, we have four kinds of recursive equations 
(I, II, III and IV in Table \ref{TableClassificationOfBCRE}). 
we also perform the same procedure shown above. Here we only show the 
identification: 
\begin{align}
{\rm I}:\qquad & 
\frac{\sigma_{m,b}^{(2s)}}{\sigma_{0,b}^{(2s)}} = \frac{-1}{\tau_m^{(2s)}},\qquad 
\frac{\sigma_{m-n,b}^{(2s)}}{\sigma_{0,b}^{(2s)}} = \frac{\tau_n^{(2s)}}{\tau_m^{(2s)}} \qquad \bigl(n=1,2,\cdots,m-1;\, 1\leq b <l\bigr),\nn\\
{\rm II}:\qquad & 
\sigma_{m,l}^{(2s)} = \frac{-1}{\tau_m^{(2s)}},\qquad 
\sigma_{m-n,l}^{(2s)} = \frac{\tau_n^{(2s)}}{\tau_m^{(2s)}} \qquad \bigl(n=1,2,\cdots,m-1\bigr),\nn\\
{\rm III}:\qquad & 
\sigma_{n,b}^{(2s)} = 0 \qquad \bigl(1<b<r;\, n=1,2,\cdots,m\bigr), \nn\\
{\rm IV}:\qquad & 
\sigma_{m-n+1,r}^{(2s)} = \tau_n^{(2s)},\qquad \bigl(n=1,2,\cdots,m\bigr), 
\end{align}
then we have a single recursion equation Eq.~\eq{JSrecursion} 
which again results in $r$ different T-systems. 
Therefore they are given by the discrete solution expression Eq.~\eq{DiscreteSolutionTauFormula}. 
However, the conditions for the indices are not straightforward in this case, 
and therefore, we show special solutions of simple cases. 

\ \\
\noindent
\underline{\bf The case of $(k,r)=(5,3)$ (i.e.~$(m,l)=(1,2)$)} 

\vspace{0.2cm}

\noindent
We found two types of solutions. First we put 
\begin{align}
\sigma_{0,1}^{(0)}: \text{ free},\qquad 
\sigma_{0,1}^{(2a)} = \omega^{\eta_a} \qquad \bigl(a=1,2\bigr),\qquad 
n\equiv n_1^{(0)},\qquad \widetilde n \equiv n_1^{(2)} = n_1^{(4)}. 
\end{align}
Then the constraints are given as 
\begin{align}
\text{1. }\quad & 
n\not \equiv  \widetilde n \quad \mod k,\qquad \eta_1=n+\widetilde n,
\qquad \eta_2 = -n-2\widetilde n\quad \mod k, \nn\\
\text{2. }\quad & 
n\equiv \widetilde n \quad \mod k,\qquad \eta_1+\eta_2 \equiv -n. 
\end{align}

\ \\
\noindent 
\underline{\bf The case of $(k,r)=(14,3)$ (i.e.~$(m,l)=(4,2)$)} 

\vspace{0.2cm}

\noindent
We found the following solutions: 
First we put 
\begin{align}
&\sigma_{n,1}^{(2a)}=0 \qquad \bigl(n=0,1,2,\cdots,m;\, 0\leq a\leq r-1\bigr),\nn\\
&n_j\equiv n_j^{(2a)} \qquad \bigl(1\leq j\leq m;\, 0\leq a\leq r-1\bigr). 
\end{align}
Then the constraints are given as 
\begin{align}
\sum_{j=1}^{m} n_j \equiv \frac{k}{2},\qquad n_j\neq k. 
\end{align}

\subsubsection{The case of $k=r(m+1)$}

In this case, we have two kinds of recursive equations 
(I and V in Table \ref{TableClassificationOfBCRE}). 
we also perform the same procedure shown above. Here we only show the 
identification: 
\begin{align}
{\rm I}:\qquad & 
\sigma_{n,b}^{(2s)} = 0 \qquad \bigl(1\leq b<r;\, n=1,2,\cdots,m+1\bigr), \nn\\
{\rm V}:\qquad & 
\sigma_{m-n+1,r}^{(2s)} = \tau_n^{(2s)},\qquad \bigl(n=1,2,\cdots,m\bigr), \qquad 
\tau_{m+1}^{(2s)} = 1, 
\end{align}
then we have a single recursion equation Eq.~\eq{JSrecursion} 
which again results in $r$ different T-systems 
(i.e.~each T-system is $U_q (A_m^{(1)})$ of $q^k=1$). 
To avoid confusion, 
we here show the boundary conditions which are slightly different from the other cases: 
\begin{align}
&T_{-n,s}(u)=T_{m+1+n,s}(u)  =0 \qquad \bigl(n \in \mathbb N,\, s\in \mathbb Z\bigr),\label{TypeIIITsystemBC1}\\
&T_{a,s+k-m-1} (u)=0\qquad \bigl(0< a <m+1, \, 0 < s < m+1\bigr), \label{TypeIIITsystemBC2}\\
&T_{a,0}(u)=T_{0,s}(u) =T_{m+1,s}(u)=1\qquad (0 \leq a \leq m+1,\, s\in \mathbb Z), \label{TypeIIITsystemBC3}
\end{align}
Therefore they are given by the discrete solution expression Eq.~\eq{DiscreteSolutionTauFormula}. 
\begin{align}
\tau_n(u;a) = (-1)^{n+1}{\rm Sym}\Bigl[n,\bigl\{\omega^{n_i^{(2a)}}\bigr\}_{i=1}^{m+1}\Bigr]\qquad 
\bigl(a=0,1,\cdots,r-1\bigr),
\end{align}
with the following conditions on the indices: 
\begin{align}
1.\qquad &n_i^{(2a)} \not\equiv n_j^{(2a)} \quad \mod k \qquad \bigl(i\neq j;\, 0\leq a\leq r-1\bigr). 
\label{TypeIIIConstraintOnIndexN1} \\
2.\qquad & 0 \equiv \sum_{j=1}^{m+1} n_j^{(2a)} + r\frac{m(m+1)}{2} \quad \mod k
\qquad \bigl(0\leq a\leq r-1\bigr).  
\label{TypeIIIConstraintOnIndexN2} \\
3.\qquad & \bigl\{n_j^{(2a)}\bigr\}_{1\leq j\leq m} = \bigl\{n_j^{(2(r-a))}\bigr\}_{1\leq j\leq m} \qquad 
\bigl(0\leq a \leq r-1\bigr). 
\label{TypeIIIConstraintOnIndexN3}
\end{align}
Note that the fractional number in Eq.~\eq{TypeIIIConstraintOnIndexN2} is always an integer. 

\section{Conclusion and discussions \label{SectionConclusionDiscussion}}

In this paper, we analyzed Stokes phenomena in the $k\times k$ isomonodromy systems with 
an arbitrary Poincar\'e index $r$, especially which correspond to the fractional-superstring 
(or parafermionic-string) multi-critical points $(\hat p,\hat q)=(1,r-1)$ in the multi-cut two-matrix models. 
Throughout the analysis, the multi-cut boundary conditions proposed in \cite{CIY2} 
turn out to be a key concept in order to uncover the underling Hirota equations (T-systems) 
and also a powerful tool in order to show explicit expressions of the Stokes multipliers in quite general class of $(k,r)$. 
Although we here focused on fractional-superstring critical points of the multi-cut two-matrix models 
(i.e.~$\gamma=r-2$), 
according to the procedure of imposing the multi-cut boundary condition discussed in Appendix \ref{SketchOfProofTheoremMCBCRE}, 
it is clear that the multi-cut BC recursion equations in Theorem \ref{TheoremMultiCutBCRecursionEquations} 
keep the universal form 
in the general critical points (i.e.~arbitrary $\gamma$) of the multi-cut two-matrix models. 

It should be noted that the basic observables in this paper, Stokes multipliers, are integration constants 
of string equations (i.e.~non-perturbative ambiguities) 
and are information completely out of the perturbative framework of string theory. Therefore, 
it is quite striking to show existence of a rigid mathematical structure like quantum integrability 
which governs these non-perturbative ambiguities of perturbative string theory. 
This fact also makes it evidential that there is 
a universal strong-string-coupling dual theory which tightly connects 
various non-critical perturbative string theories in non-perturbative regime. 

At the first sight, the multi-cut boundary condition still seems like artificial constraints 
which specialize the solutions (in string equation) originated from the matrix models. 
In particular, the physical origin of the complementary boundary conditions is still unclear. 
In the future study, we should clarify this point and uncover the physical origin of these constraints. 
There is, however, an interesting possibility to solve this problem. 
It is about a continuum description of $(l,k)$-FZZT branes. 
From the CFT approach, the boundary states are described by a superposition of $(1,1)$-FZZT branes 
with a shift of boundary cosmological constants \cite{SeSh,KOPSS}: 
\begin{align}
\Bigl|\tau;(l,k)\Bigr>_{\rm FZZT} = \sum_{m,n} \Bigl|\tau+\bigl(\frac{m}{p}+ \frac{n}{q}\bigr) i \pi ;(1,1)\Bigr>_{\rm FZZT},
\qquad \zeta =\sqrt{\mu} \cosh ( p \tau). \label{SeShRelation}
\end{align}
The description in the matrix models was also found and studied in \cite{BIR}. 
From this point of views, the general interpretation is that the $(1,1)$-FZZT branes are fundamental 
and the others are expressed by them and that this is a reason why we have a single resolvent 
operator which describe the fundamental one. 
However, the above formula also suggests the following thing: 
The shifts of the momentum in Eq.~\eq{SeShRelation} are 
essentially related to choosing different (independent) solutions of the Baker-Akhiezer system \eq{DESystem} 
on which we impose the complementary boundary conditions. 
Therefore, this implies the following possibility: 
{\em it is true that all-order perturbative amplitudes of $(l,k)$-FZZT branes 
are equivalent to those of the $(1,1)$-FZZT branes 
with the relation \eq{SeShRelation},
however they are described as non-perturbatively different objects in string theory}.%
\footnote{After publication of this paper, we were informed that it was already observed that 
the Seiberg-Shih relation \eq{SeShRelation} does not generally hold beyond the disk-amplitude level 
(i.e.~even at the perturbative level) \cite{Gesser,Atkin}. 
We would like to thank Max R. Atkin for drawing our attention to the papers. } 
That is, the above relation is accidental in perturbation theory 
and these $(l,k)$-FZZT branes have different realizations in the matrix models. 
If this consideration is true, 
then this would mean that {\em the non-perturbative completion of string theory requires 
non-perturbative realization of all the D-brane spectrum in the theory}.

Another approach to the physical meaning would be theoretical consistency of non-critical M theory, 
i.e.~the requirement that non-critical M theory can be consistently reconstructed. 
It is because we expect that the non-perturbative ambiguities should have 
definite physical meaning in the strong-coupling dual theory, 
although non-perturbative ambiguities cannot be fixed in the perturbative framework.%
\footnote{Generally speaking, we require that string theory should be completed 
by its dual theories in every theoretical limit of the backgrounds.}
An interesting clue to this approach is 
$\mathcal N=2$ QFT of Cecotti-Vafa \cite{CecottiVafa}. 
Interestingly, there appear the $k\times k$ isomonodromy systems of Poincar\'e index $r=1$ 
and therefore one can find a translation into non-critical string/M theory 
(shown in Table \ref{CecottiVafaAndNCMT}). As one can see, non-critical M theory is a simple 
generalization of $\mathcal N=2$ systems to the general Poincar\'e index $r$. 
The critical points of $r=1$ (i.e.~$\hat q=0$) in the multi-cut two-matrix models 
are lower critical points than the first critical points ($r=2$) which is the ``pure-gravity'' counterpart in the multi-cut 
two-matrix models. This situation is an analogy of the relation 
between topological series $(p,q)=(1,q)$ in minimal string theories and the other 
critical points which start with the pure-gravity $(p,q)=(2,3)$ points (i,e.~$p\geq 2$). 
In this sense, the reconstruction of the multi-critical $(r\geq 2)$ non-critical M theory is to find the multi-critical 
generalization of $\mathcal N=2$ QFTs. Also in this reconstruction, it is interesting to see 
how the quantum integrability of the Stokes multipliers play its rule. 
It is also interesting if there is a connection between 
quantum integrability of the multi-cut boundary conditions and the one of Nekrasov-Shatashvili 
\cite{NekrasovShatashvili}.

\begin{table}[t]
\begin{center}
\begin{tabular}[c]{c|c}
\hline
\small Non-critical M theory ($r>1$) & \small $\mathcal N=2$ QFTs ($r=1$) \\
\hline \hline 
\small perturbative string-theory sectors \small & local vacua $\del_x W(x)=0$ \\
\hline
\small ZZ-branes connecting string theories & \small solitons connecting local vacua \\
\hline 
\small Stokes multipliers (D-instanton fugacity) & \small Stokes multipliers (the number of solitons) \\
\hline
\end{tabular}
\end{center}
\caption{\footnotesize The dictionary of isomonodromy terminology 
between non-critical M theory and $\mathcal N=2$ QFTs. Local vacua in $\mathcal N=2$ QFTs are 
interpreted as perturbative string-theory sectors in non-critical M theory (of weak coupling limit). 
Solitons among the vacua are understood as ZZ branes among the perturbative string-theory sectors. 
Interestingly, the Stokes multipliers in $\mathcal N=2$ QFTs are interpreted as physical observables. 
Therefore, this suggests that the D-instanton chemical potentials in string theory 
should be naturally calculable as physical observables 
in non-critical M theory. }
\label{CecottiVafaAndNCMT}
\end{table}

The appearance of the quantum integrability in the Stokes multipliers of the general isomonodromy systems 
also open interesting directions. In particular, since the spectral parameter in non-critical string theory 
is understood as the spacetime coordinate \cite{fy3,MMSS}, it is interesting if 
the new spectral parameters from the T-systems are understood as the third dimension of non-critical M theory.

\vspace{1cm}
\noindent
{\bf \large Acknowledgment}  
\vspace{0.2cm}

\noindent
The authors would like to thank Jean-Emile Bourgine, Kazuyuki Furuuchi, 
Martin Guest, Chang-Shou Lin, Toshio Nakatsu, Ricardo Schiappa, 
Ryo Suzuki, Dan Tomino and Kentaroh Yoshida for useful discussions and comments. 
H.~Irie would like to thank organizers of Strings 2011 
for giving him the opportunity of presenting basic results of this paper in the Gong Show session 
in the conference. H.~Irie would like to thank people in IST for their hospitality during his visit 
while completing this work. 
C.-T.~Chan and H.~Irie are supported in part by National Science Council of Taiwan under the contract 
No.~99-2112-M-029-001-MY3 (C.-T.~Chan) and No.~100-2119-M-007-001 (H.~Irie). 
The authors are also supported in part by Taiwan String Theory Focus Group 
in National Center for Theoretical Science under NSC No.~99-2119-M-007-001 and No.~100-2119-M-002-001.

\appendix

\section{Sketch of a proof for Theorem \ref{TheoremMultiCutBCRecursionEquations}
\label{SketchOfProofTheoremMCBCRE}}

Here we show a brief sketch of a proof for 
Theorem \ref{TheoremMultiCutBCRecursionEquations}. 
The procedure is the following: 
\begin{itemize}
\item [1.] Introduce a division the vectors (Eq.~\eq{MCBCofXforGeneralKWithR3} for $r=3$) 
and the equation Eq.~\eq{MCBCRecursionVectorNotation} of $n=0$ 
according to modulo $r$. 
\item [2.] Write down all the equation Eq.~\eq{MCBCRecursionVectorNotation} of $n=0$ 
according to the modulo-$r$ division. 
\item [3.] Obtain a closed set of equations, by using proper shift rules. 
\item [4.] Solve them and obtain equations with non-zero $x_n^{(2s)}$. With identifying the 
sequence $\bigl\{y_n^{(2s)}\bigr\}$ with non-zero $x_n^{(2s)}$, these equations 
result in Eq.~\eq{GeneralFormulaForTheMultiCutBCRecursionEquation}. 
The shift rules of the equation then generate all the equations of Eq.~\eq{MCBCRecursionVectorNotation}. 
\end{itemize}
Here we demonstrate this procedure in the case of $(k,r)=(6\widetilde m+1,3)$. 

\paragraph{Step 1} Here we introduce a modulo-$r$ division of the vectors 
Eq.~\eq{MCBCofXforGeneralKWithR3} and Eq.~\eq{MCBCRecursionVectorNotation} for $n=0$. 
The division is given in the following way:
\begin{align}
&
\left. \left.
\begin{array}{ccccccccc}
&X^{(6)} & = & \Bigl( & 4^\triangle, & \times, & \cdots, & \times, &
\bigl(3\widetilde m+1\bigr)^\triangle \cr
S_0^{(\rm sym)}: & \downarrow  & & & \downarrow &  & \cdots & &  \downarrow \cr
& X^{(0)} & = &\Bigl( & 4^\triangle, & \times, & \cdots, & \times, & 
\bigl(3\widetilde m+1\bigr)^\triangle 
\end{array}
\right|  \,\, \right\}:
\begin{minipage}{4cm}
There are only non-zero components, $i^\triangle$. 
\end{minipage}
\nn\\
&\qquad \qquad 
\underbrace{
\left|
\begin{array}{ccc|ccc|ccc|cccc}
\times, & \times, & \bigl(3\widetilde m+4\bigr)^\triangle &%
 -,& \cdots, & -&%
k-2,& k-1,& k &%
1,&2,&3 & \Bigr) \cr
\downarrow &\downarrow &\downarrow  &%
\downarrow &  \cdots & \downarrow &%
\downarrow & \downarrow & \downarrow &%
\downarrow &\downarrow &\downarrow  & \cr
-,&-,&- &%
 -,& \cdots, & - &%
 k-2,& k-1,& k &%
 1^\triangle, & \times, & \times & \Bigr)
\end{array}
\right.
}_{\text{\normalsize The number of the components is a multiple of $r$.}}
\label{DivisionOfRegionsInVectorXR3}
\end{align}
Here the components $(1,2,3)$ are moved to the next to the component ``$k$''. 
The meaning of the arrow is
\begin{align}
\begin{array}{cc}
& i \cr
S_0^{(\rm sym)}: & \downarrow \cr
& i
\end{array}
\quad 
\Leftrightarrow 
\quad 
x_i^{(0)} = x_i^{(6)} + \sum_{j} s_{0,i,j^{\triangle}}^{(\rm sym)}\, x_{j^{\triangle}}^{(6)}.
\end{align}
Here ``$\triangle$'' of $x_{j^\triangle}^{(6)}$ means that the $x_{j^\triangle}^{(6)}\neq 0$. 
One can always prove that the contributions with Stokes multipliers are always accompanied 
with the non-zero component $x_{j^\triangle}^{(6n)}$. 
Then we first divide them into a part which consists only of non-zero components $i^\triangle$ 
and the other part, in which the number of components is a multiple of $r$. 
We introduce modulo-$r$ division in the latter part 
so that the right-hand side of $k$ should be boundary of the division, 
``$\cdots,k | 1, 2,3)$''. In this way, the equations below become simpler. 

\paragraph{Step 2} We write all the equation Eq.~\eq{MCBCRecursionVectorNotation} for $n=0$:
\begin{align}
{\rm I.}\quad &\left[
\begin{array}{ccccc}
4^\triangle, & \times, & \cdots, & \times, &
\bigl(3\widetilde m+1\bigr)^\triangle \cr
4^\triangle, & \times, & \cdots, & \times, &
\bigl(3\widetilde m+1\bigr)^\triangle
\end{array}
\right]\leftrightarrow
\left\{
\begin{array}{rr}
\ds x_{(3n+1)^\triangle}^{(0)} = x_{(3n+1)^\triangle}^{(6)}, & \bigl(n=1,2,\cdots,\widetilde m\bigr)
\end{array}
\right.,\nn\\
{\rm II.}\quad &\left[
\begin{array}{ccc}
1,&2,&3 \cr
1^\triangle,&\times,&\times 
\end{array}
\right]\leftrightarrow
\left\{
\begin{array}{r}
\ds 0 = x_3^{(6)} + \theta_1^{(0)} x_{4^\triangle}^{(0)} \cr
\ds 0 = x_2^{(6)} + \theta_2^{(0)} x_{4^\triangle}^{(0)} \cr
\ds x_{1^\triangle}^{(0)} = x_1^{(6)} + \theta_3^{(0)} x_{4^\triangle}^{(0)}
\end{array}
\right.,\nn\\
{\rm III.}\quad &\left[
\begin{array}{ccc}
k-3n+1,&k-3n+2,&k-3n + 3 \cr
k-3n+1,&k-3n+2,&k-3n + 3
\end{array}
\right] \quad \bigl(n=1,2,\cdots,\widetilde m-1\bigr)\nn\\
&\qquad \leftrightarrow
\left\{
\begin{array}{c}
\ds x_{k-3n+a}^{(0)} = x_{k-3n+a}^{(6)} 
+ \theta_{6n+1-a}^{\bigl(6(n-1)\bigr)} x_{(3n+1)^\triangle}^{(0)} 
+ \theta_{6n+4-a}^{(6n)} x_{(3n+4)^\triangle}^{(0)},\quad  \bigl(a=1,2,3\bigr), 
\end{array}
\right.\nn\\
{\rm IV.}\quad &\left[
\begin{array}{ccc}
\times,&\times,&\bigl(3\widetilde m+4\bigr)^\triangle \cr
3\widetilde m+2,&3\widetilde m+3,&3\widetilde m+4
\end{array}
\right] =
\left[
\begin{array}{ccc}
\times,&\times,&\bigl(k-3\widetilde m+3\bigr)^\triangle \cr
k-3\widetilde m+1&k-3\widetilde m+2,&k-3\widetilde m+3
\end{array}
\right]
\nn\\
&\qquad \leftrightarrow
\left\{
\begin{array}{l}
\ds x_{3\widetilde m + 4}^{(0)} 
= x_{(3\widetilde m + 4)^\triangle}^{(6)} + \theta_{6\widetilde m-2}^{\bigl(6(\widetilde m-1)\bigr)} x_{(3\widetilde m+1)^\triangle}^{(0)} \cr
\ds x_{3\widetilde m + 3}^{(0)} 
= \theta_{6\widetilde m-1}^{\bigl(6(\widetilde m-1)\bigr)} 
x_{(3\widetilde m+1)^\triangle}^{(0)} \cr
\ds x_{3\widetilde m + 2}^{(0)} 
=  \theta_{6\widetilde m}^{\bigl(6(\widetilde m-1)\bigr)} x_{(3\widetilde m+1)^\triangle}^{(0)}
\end{array}
\right..
\end{align}

\paragraph{Step 3} We make a shift rule to obtain equations among $\bigl\{x_1^{(6n)},x_2^{(6n)},x_3^{(6n)}\bigr\}_{n\in \mathbb Z}$. 
\begin{align}
{\rm III.}\quad & 
\left\{
\begin{array}{c}
\ds x_{k+a}^{(6n)} = x_{k+a}^{(6(n+1))} 
+ \theta_{6n+1-a}^{\bigl(6(2n-1)\bigr)} x_{(1+6n)^\triangle}^{(6n)} 
+ \theta_{6n+4-a}^{\bigl(6(2n)\bigr)} x_{(4+6n)^\triangle}^{(6n)},
\quad \bigl(a=1,2,3\bigr)
\end{array}
\right. \nn\\
{\rm IV.}\quad & \left\{
\begin{array}{c}
\ds x_{3}^{(6\widetilde m)} 
= x_{3^\triangle}^{(6(\widetilde m+1))} + \theta_{6\widetilde m-2}^{\bigl(6(2\widetilde m-1)\bigr)} x_{(6\widetilde m+1)^\triangle}^{(6\widetilde m)} \cr
\ds x_{2}^{(6\widetilde m)} 
= \theta_{6\widetilde m-1}^{\bigl(6(2\widetilde m-1)\bigr)} 
x_{(6\widetilde m+1)^\triangle}^{(6\widetilde m)} \cr
\ds x_{1}^{(6\widetilde m)} 
=  \theta_{6\widetilde m}^{\bigl(6(2\widetilde m-1)\bigr)} x_{(6\widetilde m+1)^\triangle}^{(6\widetilde m)}
\end{array}
\right.. 
\end{align}

\paragraph{Step 4} By solving these equations, we obtain equations for non-zero 
components of $\bigl\{x_i^{(6n)}\bigr\}_{i,n\in \mathbb Z}$:
\begin{align}
\left\{
\begin{array}{c}
\ds 0
= x_{3^\triangle}^{(6(\widetilde m+1))} 
+ \sum_{n=1}^{\widetilde m}\theta_{6n-2}^{\bigl(6(2n-1)\bigr)} x_{(1+6n)^\triangle}^{(6n)} 
+ \sum_{n=0}^{\widetilde m-1}\theta_{6n+1}^{\bigl(6(2n)\bigr)} x_{(4+6n)^\triangle}^{(6n)}, \cr
\ds 0
= 
\sum_{n=1}^{\widetilde m}\theta_{6n-1}^{\bigl(6(2n-1)\bigr)} x_{(6n+1)^\triangle}^{(6n)} 
+ \sum_{n=0}^{\widetilde m-1}\theta_{6n+2}^{\bigl(6(2n)\bigr)} x_{(6n+4)^\triangle}^{(6n)}, \cr
\ds x_{1^\triangle}^{(0)} 
=  
\sum_{n=1}^{\widetilde m}\theta_{6n}^{\bigl(6(2n-1)\bigr)} x_{(1+6n)^\triangle}^{(6n)} 
+ \sum_{n=0}^{\widetilde m-1}\theta_{6n+3}^{\bigl(6(2n)\bigr)} x_{(4+6n)^\triangle}^{(6n)}.
\end{array}
\right.
\end{align}
By introducing the following sequence $\bigl\{y_n^{(0)}\bigr\}_{n \in\mathbb Z}$, 
\begin{align}
y_{n}^{(0)} \equiv x_{(3n+1)^\triangle}^{(6n)}
\Bigl(=x_{(3n+1)^\triangle}^{(6(n-1))}=\cdots 
=x_{(3n+1)^\triangle}^{(6(n-\widetilde m))}\Bigr),
\end{align}
we obtain the equations of the form: 
\begin{align}
\mathcal F_1 \Bigl[\bigl\{y_n^{(0)}\bigr\}_{n \in\mathbb Z}\Bigr] 
&=  y_{2\widetilde m+1}^{(0)} + \sum_{n=1}^{2\widetilde m} \theta_{3(n-1)+1}^{(6(n-1))}\, y_n^{(0)} = 0, \nn\\
\mathcal F_2 \Bigl[\bigl\{y_n^{(0)}\bigr\}_{n \in\mathbb Z}\Bigr] 
&=  \sum_{n=1}^{2\widetilde m} \theta_{3(n-1)+2}^{(6(n-1))}\, y_n^{(0)} = 0, \nn\\
\mathcal F_3 \Bigl[\bigl\{y_n^{(0)}\bigr\}_{n \in\mathbb Z}\Bigr] 
&=  -y_{0}^{(0)} + \sum_{n=1}^{2\widetilde m} \theta_{3(n-1)+3}^{(6(n-1))}\, y_n^{(0)} = 0. 
\end{align}
By taking into account Eq.~\eq{Theta0k}, 
we obtain Eq.~\eq{GeneralFormulaForTheMultiCutBCRecursionEquation} with $s=0$. 
Therefore, all the other equations in Eq.~\eq{GeneralFormulaForTheMultiCutBCRecursionEquation} are obtained by the $2s$ shift rule. 
Note that $y_n^{(2s)}$ is defined as 
\begin{align}
y_n^{(2s)} = x^{(6n+2s)}_{(3n+s+1)^\triangle}. 
\end{align}
In this sense, we use the notation 
of $k =rm+l$ $(1\leq l\leq r)$ in the following discussions. 
Finally we mention the extension of this procedure to all the $(k,r)$ cases. 
We note is that we have used a specialty of $r=3$, 
i.e.~$m_1=1$ (See Eq.~\eq{ProfileShiftRule}). Because of this specialty, 
the division procedure Eq.~\eq{DivisionOfRegionsInVectorXR3} is simpler 
and the proof is not so complicated.%
\footnote{See \cite{CIY2} for $r=2$, which still looks complicated.} 
In the general cases, however, 
we can re-organize the components of the vector $X^{(0)}$ as
\begin{align}
X^{(0)} \rightarrow (x_1^{(0)},x_{1+r m_1}^{(0)},x_{1+2 r m_1}^{(0)},\cdots).
\end{align}
In this way, we can perform the same procedure in general $r$ and can 
show Eq.~\eq{GeneralFormulaForTheMultiCutBCRecursionEquation} for general 
$(k,r)$.

\end{document}